%% file: paper.tex
\documentclass[fleqn,usenatbib]{mnras}

\usepackage[T1]{fontenc}

\DeclareRobustCommand{\VAN}[3]{#2}
\let\VANthebibliography\thebibliography
\def\thebibliography{\DeclareRobustCommand{\VAN}[3]{##3}\VANthebibliography}

\usepackage{graphicx}	% Including figure files
\usepackage{amsmath}	% Advanced maths commands
\usepackage{amssymb}	% Extra maths symbols

\usepackage{newtxtext,newtxmath}

\makeatletter
\DeclareRobustCommand{\ion}[2]{%
  \text{#1\,\check@mathfonts\fontsize\sf@size\z@\selectfont #2}%
}
\makeatother

\newcommand{\met}{\ensuremath{12+\log_{10}(\mathrm{O/H})}}
\newcommand{\oh}{\ensuremath{\log_{10}(\mathrm{O/H})}}
\newcommand{\ha}{\textup{H\ensuremath{\upalpha}}}
\newcommand{\hb}{\textup{H\ensuremath{\upbeta}}}
\newcommand{\oiii}{[\ion{O}{III}]\ensuremath{\lambda5006}}
\newcommand{\oiiijoin}{[\ion{O}{III}]\ensuremath{\lambda\lambda4958{,}5006}}
\newcommand{\sii}{[\ion{S}{II}]\ensuremath{\lambda\lambda6716{,}30}}
\newcommand{\nii}{[\ion{N}{II}]\ensuremath{\lambda6583}}
\newcommand{\niijoin}{[\ion{N}{II}]\ensuremath{\lambda\lambda6548{,}83}}
\newcommand{\hii}{\ion{H}{II}} % no trailing space
\newcommand{\ngal}{12}

\title[PHANGS Metallicity Maps]{The Two-Dimensional Metallicity Distribution and Mixing Scales of Nearby Galaxies}

\author[T.~G.~Williams et al.]{Thomas~G.~Williams,$^{1}$\thanks{E-mail: williams@mpia.de (TGW)}
Kathryn Kreckel,$^{2}$
Francesco Belfiore,$^{3}$
Brent Groves,$^{4}$
Karin Sandstrom,$^{5}$
\newauthor
Francesco Santoro,$^{1}$
Guillermo A. Blanc,$^{6,7}$
Frank Bigiel,$^{8}$
M{\'e}d{\'e}ric Boquien,$^{9}$
M{\'e}lanie Chevance,$^{10}$
\newauthor
Enrico Congiu,$^{7}$
Eric Emsellem,$^{11,12}$
Simon C. O. Glover,$^{13}$
Kathryn Grasha,$^{14}$
\newauthor
Ralf S.\ Klessen,$^{13,15}$
Eric Koch,$^{16}$
J.~M.~Diederik Kruijssen,$^{10}$
Adam K. Leroy,$^{17}$
Daizhong Liu,$^{18}$
\newauthor
Sharon Meidt,$^{19}$
Hsi-An Pan,$^{1,20}$
Miguel Querejeta,$^{21}$
Erik Rosolowsky,$^{22}$
Toshiki Saito,$^{23,24}$
\newauthor
Patricia S{\'a}nchez-Bl{\'a}zquez,$^{25}$
Eva Schinnerer,$^{1}$
Andreas Schruba,$^{18}$
and Elizabeth~J.~Watkins$^{2}$
\\
$^{1}$Max-Planck-Institut f\"{u}r Astronomie, K\"{o}nigstuhl 17, D-69117, Heidelberg, Germany\\
$^{2}$Astronomisches Rechen-Institut, Zentrum f\"{u}r Astronomie der Universit\"{a}t Heidelberg, M\"{o}nchhofstra\ss e 12-14, D-69120 Heidelberg, Germany\\
$^{3}$INAF – Osservatorio Astrofisico di Arcetri, Largo E. Fermi 5, I-50157, Firenze, Italy\\
$^{4}$International Centre for Radio Astronomy Research, University of Western Australia, 7 Fairway, Crawley, 6009, WA, Australia\\
$^{5}$Center for Astrophysics and Space Sciences, Department of Physics, University of California, San Diego, 9500 Gilman Dr., La Jolla, CA 92093, USA\\
$^{6}$The Observatories of the Carnegie Institution for Science, 813 Santa Barbara Street, Pasadena, CA 91101\\
$^{7}$Departamento de Astronom{\'i}a, Universidad de Chile, Camino del Observatorio 1515, Las Condes, Santiago, Chile\\
$^{8}$Argelander-Institut f{\"u}r Astronomie, Universit{\"a}t Bonn, Auf dem H{\"u}gel 71, 53121 Bonn, Germany\\
$^{9}$Universidad de Antofagasta, Centro de Astronom{\'i}a, Avenida Angamos 601, Antofagasta 1270300, Chile \\
$^{10}$Astronomisches Rechen-Institut, Zentrum f\"{u}r Astronomie der Universit\"{a}t Heidelberg, M\"{o}nchhofstra\ss e 12-14, D-69120 Heidelberg, Germany\\
$^{11}$European Southern Observatory, Karl-Schwarzschild-Str. 2, 85748 Garching, Germany\\ 
$^{12}$Universit{\'e} Lyon 1, ENS de Lyon, CNRS, Centre de Recherche Astrophysique de Lyon UMR5574, 69230 Saint-Genis-Laval, France\\
$^{13}$Universit{\"a}t Heidelberg, Zentrum f{\"u}r Astronomie Heidelberg, Institut f{\"u}r Theoretische Astrophysik, Albert-Ueberle-Str. 2, 69120 Heidelberg, Germany\\
$^{14}$Research School of Astronomy and Astrophysics, Australian National University, Weston Creek, ACT 2611, Australia\\
$^{15}$Universit{\"a}t Heidelberg, Interdisziplin{\"a}re Zentrum f{\"u}r Wissenschaftliches Rechnen, Im Neuenheimer Feld 205, 69120 Heidelberg, Germany\\
$^{16}$Center for Astrophysics | Harvard \& Smithsonian, 60 Garden St., 02138 Cambridge, MA, USA\\
$^{17}$Department of Astronomy, The Ohio State University, 140 West 18th Avenue, Columbus, Ohio 43210, USA\\
$^{18}$Max-Planck-Institut f{\"u}r extraterrestrische Physik, Giessenbachstra{\ss}e~1, D-85748 Garching, Germany\\
$^{19}$Sterrenkundig Observatorium, Universiteit Gent, Krijgslaan 281 S9, B-9000 Gent, Belgium\\
$^{20}$Department of Physics, Tamkang University, Tamsui Dist., New Taipei City 251301, Taiwan\\
$^{21}$Observatorio Astron{\'o}mico Nacional (IGN), C/Alfonso XII 3, Madrid E-28014, Spain\\
$^{22}$Dept. of Physics, University of Alberta, Edmonton, Alberta, Canada T6G 2E1\\
$^{23}$Department of Physics, General Studies, College of Engineering, Nihon University, 1 Nakagawara, Tokusada, Tamuramachi, Koriyama,\\ \hspace{0.5cm} Fukushima, 963-8642, Japan\\
$^{24}$National Astronomical Observatory of Japan, 221-1 Osawa, Mitaka, Tokyo, 181-8588, Japan\\
$^{25}$Departamento de F{\'i}sica de la Tierra y Astrof{\'i}sica, Facultad de CC F{\'i}sicas, Universidad Complutense de Madrid, 28040, Madrid, Spain\\
}

\date{Accepted XXX. Received YYY; in original form ZZZ}

\pubyear{2021}

\begin{document}
\label{firstpage}
\pagerange{\pageref{firstpage}--\pageref{lastpage}}
\maketitle
\clearpage

% Abstract of the paper - 249/250 words
\begin{abstract}
Understanding the spatial distribution of metals within galaxies allows us to study the processes of  chemical enrichment and mixing in the interstellar medium (ISM). In this work, we map the two-dimensional distribution of metals using a Gaussian Process Regression (GPR) for 19 star-forming galaxies observed with the Very Large Telescope/\linebreak[0]{}Multi Unit
Spectroscopic Explorer (VLT--MUSE) as part of the PHANGS--MUSE survey. We find that \ngal\ of our 19 galaxies show significant two-dimensional metallicity variation. Those without significant variations typically have fewer metallicity measurements, indicating this is due to the dearth of \hii\ regions in these galaxies, rather than a lack of higher-order variation. After subtracting a linear radial gradient, we see no enrichment in the spiral arms versus the disc. We measure the 50~per~cent correlation scale from the two-point correlation function of these radially-subtracted maps, finding it to typically be an order of magnitude smaller than the fitted GPR kernel scale length. We study the dependence of the two-point correlation scale length with a number of global galaxy properties. We find no relationship between the 50~per~cent correlation scale and the overall gas turbulence, in tension with existing theoretical models. We also find more actively star forming galaxies, and earlier type galaxies have a larger 50~per~cent correlation scale. The size and stellar mass surface density do not appear to correlate with the 50~per~cent correlation scale, indicating that perhaps the evolutionary state of the galaxy and its current star formation activity is the strongest indicator of the homogeneity of the metal distribution.
\end{abstract}

\begin{keywords}
ISM: abundances -- ISM: evolution -- galaxies: ISM -- galaxies: general
\end{keywords}

%%%%%%%%%%%%%%%%%%%%%%%%%%%%%%%%%%%%%%%%%%%%%%%%%%

%%%%%%%%%%%%%%%%% BODY OF PAPER %%%%%%%%%%%%%%%%%%

\section{Introduction}

\input{galaxy_params}

The chemical composition of the interstellar medium (ISM) plays a fundamental role in galaxy evolution. Within galaxies, stars enrich the surrounding ISM, causing chemical changes. On large scales, `inside-out' galaxy growth \citep{1999Boissier} naturally leads to a negative gradient in the gas-phase chemical composition (hereafter `metallicity') of a galaxy \citep[e.g.][]{1971Searle, 1992MartinRoy, 2019Belfiore}. On small scales, however, individual star-forming regions (the sites of individual enrichment events) are not independent and isolated, but are embedded within dynamically active galaxies. The dynamical conditions of galaxies can act to mix metals from different star-forming regions, leading to an increase in homogeneity between neighbouring regions. For instance, interstellar turbulence \citep[e.g.][]{2003KlessenLin,2018KrumholzTing}, and gravitational instability \citep[e.g.][]{2012YangKrumholz} can drive mixing and increase homogeneity. Conversely, numerical simulations predict azimuthal variations driven by spiral arms and bars \citep[e.g.][]{2013DiMatteo, 2018Fragkoudi, 2020Fragkoudi}. Recently, these variations have been observed in nearby galaxies \citep[e.g.][]{2017Ho,2019Kreckel,2020SanchezMenguiano, 2021Li}, and an important measurement of the homogeneity is a `chemical mixing scale,' which describes the spatial scales over which neighbouring regions have highly correlated metallicity values \citep[e.g.][]{2018KrumholzTing,2020Kreckel,2021Li}. As metallicity variations are driven by non-axisymmetric features, our traditional one-dimensional view of the distribution of metals is insufficient to study these phenomena. We therefore require high-quality, two-dimensional maps of the metallicity for entire galaxies.

High angular resolution is required to accurately map the metallicity in galaxies, due to contamination from gas emission lines in many density regimes, where metallicity prescriptions may not be valid. In particular, we see line emission arising from diffuse ionised gas (DIG), which typically has lower density and higher temperatures than the more dense \hii\ regions \citep{2009Haffner, 2021Belfiore}. The DIG also appears to exist in two regimes -- the DIG around \hii\ regions has line ratios consistent with that of the \hii\ regions themselves, but DIG further away from these regions is dominated by ionisation from very different sources \citep{2021Belfiore}. Whether the DIG line emission can be used to infer metallicity is unclear, although attempts have been made to measure metallicities within the DIG \citep{2019Kumari}, or remove the DIG contribution from the emission \citep{2016Kaplan,2019Poetrodjojo,2019ValeAsari}. Until a robust method for measuring metallicities is developed that accounts for variations in the ionising spectrum, our best effort is to measure metallicities in \hii\ regions where metallicity calibrations are valid, and then interpolate between them.

With the recent advent of optical integral field units (IFUs), such as the Multi Unit Spectroscopic Explorer \citep[MUSE;][]{2010Bacon} on the Very Large Telescope (VLT), it is now possible to identify and characterise thousands of \hii\ regions across the faces of nearby galaxy discs \citep[e.g.][]{2011RosalesOrtega, 2015Sanchez, 2017Ho, 2018Poetrodjojo, 2018Sarzi, 2019ErrozFerrer, 2019Gadotti, 2019Ho, 2020Kreckel}. This gives us access to the chemical conditions of individual \hii\ regions,  allowing us to build a picture of the overall metallicity distribution. However, to move beyond simple radial gradients, we require hundreds or even thousands of metallicity measurements across a galactic disc \citep[e.g.][]{2019Clark, 2020Kreckel}. Given that \hii\ regions are not located uniformly across the disc of a galaxy, and can also be sparsely distributed \citep{2021Santoro}, we require advanced statistical techniques to model the underlying metallicity distribution of the galactic disc.

Producing a `filled-in' map for metallicities that are traditionally sparsely measured in observations has a variety of uses. In particular, a complete map is particularly powerful when combining with lower resolution, but spatially complete, data, for example {\it Herschel} IR photometry to study the dust-to-metals ratio \citep{2021Chiang} or dust mass absorption coefficient \citep{2019Clark}. These maps can also be used in conjunction with the CO maps from the Atacama Large Millimeter/\linebreak[0]{}submillimeter Array (ALMA) of the same galaxies \citep{2021Leroy}, to study, e.g., the metallicity dependence of molecular gas properties. Simple nearest-neighbour or bilinear interpolation performs poorly over the sparse, irregular sampling that metallicity measurements typically have, and so a more unbiased interpolation method is required.

In this work, we apply a Gaussian Process Regression (GPR) technique \citep[for a general introduction to this technique, see][]{2006RasmussenWilliams} to map the smooth, higher-order (i.e.\ the residual after subtracting a radial gradient) metallicity variation in a number of nearby galaxies. \cite{2019GonzalezGaitan} have recently applied a similar technique to that which we use in this work to a number of IFU cubes, finding significantly improved results over the na{\"i}ve nearest neighbour interpolation approach. The galaxies we analyse have all been mapped as part of the Physics at High Angular resolution in Nearby GalaxieS\footnote{\url{http://www.phangs.org}} (PHANGS) survey. We use data from the PHANGS--MUSE survey \citep{2021Emsellem}, which observed the star-forming discs of a sample of 19 nearby, main sequence galaxies at homogeneous sensitivity and resolution. With this large set of galaxies, we can search for trends in the typical ISM mixing scales as a function of various galaxy parameters. With the high spatial resolution that MUSE achieves within these galaxies (${\sim}100$~pc), we can also search for variations in metal enrichment within particular galactic environments (e.g.\ the centres of galaxies, or their spiral arms).

The layout of this paper is as follows. In Section~\ref{sec:data}, we present an overview of the data used in this study. In Section~\ref{sec:gpr}, we describe our procedure for modelling the two-dimensional metallicity distribution to obtain kernel length scales over which metals are homogeneously mixed, and compare these kernel length scales to correlation scales measured from a two-point correlation function. We search for environmental dependence on metallicity enrichment in Section~\ref{sec:environment}. In Section~\ref{sec:global_correlations}, we investigate the relationship between the two-point correlation scales with various galaxy parameters. Finally, we conclude in Section~\ref{sec:conclusions}. These maps are made publicly available at the links provided in the Data Availability Section.

\section{Data}\label{sec:data}

\begin{figure*}
\includegraphics[width=1.75\columnwidth]{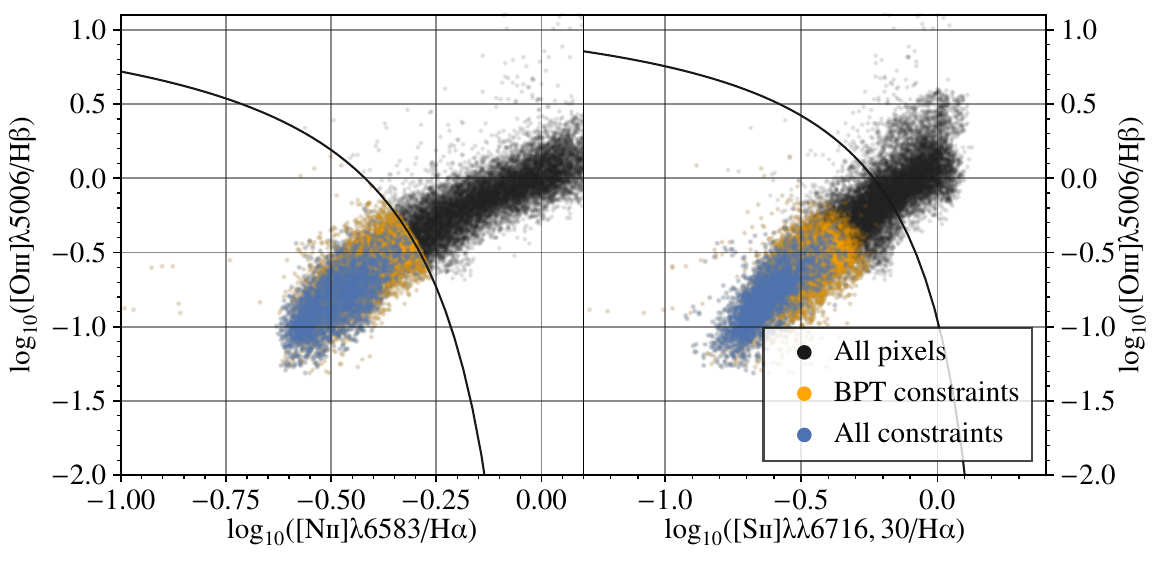}
\caption{BPT diagnostics for pixel metallicity measurements NGC~3627. In each case, the star-formation demarcation line is shown in black, with values below this line being consistent with photoionisation. {\it Left:} \citet{2003Kauffmann} diagnostic in the $\oiii/\nii$ plane. {\it Right:} \citet{2001Kewley} diagnostic in the $\oiii/\sii$ plane. In each case, we show all pixels in the convolved NGC~3627 map  as a black cloud, those that satisfy both BPT constraints (i.e. are below both black lines) in yellow, and those that satisfy all our data constraints  ($\mathrm{S/N} > 5$, velocity dispersion $< 100$, within morphologically defined \hii\ region, see Sect.~\ref{sec:data_prep};  i.e.\ those that form our final pixel map) in blue.} 
\label{fig:ngc3627_bpt}
\end{figure*}

In this study, we make use of MUSE IFU data for 19 nearby galaxies taken as part of the PHANGS--MUSE survey \citep{2021Emsellem}. Full details of the data reduction and processing are given in that work, and so we give only a short summary here. The data reduction is performed using standard MUSE recipes (e.g.\ wavelength and flux calibration, cosmic ray rejection, mosaicking) developed by the MUSE consortium and carried out by the MUSE data reduction pipeline \citep{2020Weilbacher}, available to the community via the {\tt esorex} package \citep{esorex} and accessed using a {\sc python} wrapper {\tt pymusepipe}\footnote{\url{https://github.com/emsellem/pymusepipe}}. These reduced products are then processed through a data analysis pipeline (DAP\footnote{\url{https://gitlab.com/francbelf/ifu-pipeline}}), which is run in three stages: firstly, stellar kinematics are measured (the stellar velocity and higher-order moments). Next, the properties of stellar populations are estimated (e.g.\ stellar age, mass, and metallicity). Both of these stages are performed on Voronoi binned data to a stellar continuum signal-to-noise ratio ($\mathrm{S/N}$) of~$35$, to maximise reliability. The fit is performed via \textsc{ppxf} \citep{2004CappellariEmsellem,2017Cappellari}, and makes use of E-MILES \citep{2016Vazdekis} single stellar population models of eight ages ($0.03{-}14$~Gyr, logarithmically spaced in steps of $0.22$~dex) and four metallicities ($\mathrm{[Z/H]} = [-1.5, -0.35, 0.06, 0.4]$). Only the wavelength range $4850{-}7000$~\AA\ is used in the fit (of the full $4750{-}9350$~\AA\ range), in order to avoid strong sky residuals in the reddest part of the MUSE spectral range. Finally, for individual spaxels the properties of emission lines are measured (fluxes and kinematics), via a simultaneous fit of continuum and emission lines also performed via \textsc{ppxf}. The final DAP products are 2D maps of the intensity, velocity, and higher order moments of these emission lines (and associated uncertainties), as well as the properties of the stellar populations.

We make use of the DAP products produced as part of the first public data release, and use the `optimally convolved' (COPT) products. Because each galaxy is observed with multiple MUSE pointings and mosaicked together, there are variations in seeing across the complete mosaic. The COPT mosaic accounts for this, by convolving each individual pointing to a common coarsest angular resolution both in wavelength, and across the entire galaxy before the spectral fitting is carried out. At this point, the shape of the point-spread function (PSF) is convolved from a Moffat to a Gaussian, which allows us to immediately convolve to other Gaussian resolutions via Gaussian convolution kernels in the following subsection. In Table~\ref{tab:galaxy_params} we present the global properties of our sample.

\subsection{Convolution}\label{sec:convolution}

Although the COPT DAP products represent a homogeneous angular resolution across the galaxy (and wavelength range), with the different distances and seeing conditions for the 19 galaxies, this does not represent a homogeneous spatial scale \emph{across} our sample of galaxies. We therefore convolve each intensity and (intensity-weighted) velocity dispersion map, along with associated error maps, to a fixed spatial resolution. We choose a common spatial resolution of $120$~pc (and pixel size of $60$~pc), which is advantageous for two reasons: firstly, the native PHANGS--MUSE resolution is $25{-}70$~pc, so we can incude all data at $120$~pc resolution. Secondly, at this resolution, a typical \hii\ region will be unresolved \citep{2009Hunt, 2019Kreckel, 2021Barnes, 2021Santoro}. As metallicity diagnostics are calibrated on entire \hii\ regions, it should be applied to pixels that contain a complete \hii\ region, and this pixel size ensures that. Given the typical region separation length of $100{-}300$~pc \citep{2020Chevance}, we expect each pixel to correspond to approximately a single \hii\ region.

For each galaxy, we calculate a Gaussian convolution kernel from the COPT beam to our fixed spatial resolution, where the FWHM of this kernel is given by
\begin{equation}
    {\rm FWHM_{conv}} = \sqrt{{\rm FWHM_{120\,pc}}^2 - {\rm FWHM_{COPT}}^2}.
\end{equation}
We convolve our intensity and (intensity-weighted) velocity dispersions with a PSF of ${\rm FWHM_{conv}}$. For the error maps, due to earlier convolution to produce the COPT maps, values are covariant with each other. We account for this by estimating the `per-pixel' noise that results in the measured error after smoothing. Taking a Gaussian convolution in two-dimensions, the relationship between the measured variance and the per-pixel variance is
\begin{equation}
    \sigma_{\rm measured}^2 = \sigma_{\rm pix}^2 \sum^{\infty}_{j=-\infty} \sum^{\infty}_{i=-\infty} \left(\frac{1}{2\pi\times{w_{\rm COPT}}^2} \mathrm{e}^\frac{-(i^2+j^2)}{2\pi\times{w_{\rm COPT}}^2}\right)^2.
\end{equation}
$i$ and $j$ are the pixels to be summed over, $\sigma_{\rm measured}$ and $\sigma_{\rm pix}$ are the measured and per-pixel error maps, respectively, and ${w_{\rm COPT}}$ is the standard deviation (Gaussian width, in pixels) of the COPT PSF, where we use $w$ instead of the usual $\sigma$ in the Gaussian equation to avoid confusion between this and the error maps. The summation approximates to
\begin{equation}
    \sigma_{\rm measured}^2 \simeq \sigma_{\rm pix}^2 \frac{1}{4\pi\times{w_{\rm COPT}}^2},
\end{equation}
and so rearranging we arrive at
\begin{equation}
    \sigma_{\rm pix} \simeq \sigma_{\rm measured} \times 2\sqrt{\pi} \times {w_{\rm COPT}}.
\end{equation}
The final error map is then
\begin{equation}
    \sigma_{\rm conv} = \sqrt{\sigma_{\rm pix}^2 \ast {\rm PSF_{120\,pc}^2}},
\end{equation}
where $\ast$ indicates a convolution and ${\rm PSF_{120\,pc}}$ is a Gaussian kernel with FWHM=120~pc. Following this, we Nyquist sample our maps to have two pixels across the ${\rm FWHM}_{120 \rm pc}$, and remove any pixels within one PSF of the map edge (i.e.\ a 2-pixel border), to avoid convolution artifacts. All reprojections are performed using \textsc{python}'s surface brightness conserving \texttt{reproject} algorithm (as the MUSE maps are presented in units of surface brightness). 

\subsection{Pixel-by-Pixel Metallicity Calculation}\label{sec:metallicity_calculation}

\subsubsection{Data Preparation}\label{sec:data_prep}

We calculated a gas-phase metallicity, \met, using our convolved line maps (we will use this oxygen abundance synonymously with metallicity throughout this work). We do not correct for Milky Way (Galactic) extinction, as this is already performed in the MUSE DAP. We correct fluxes for internal extinction. The internal extinction is calculated via the Balmer decrement,
\begin{equation}\label{eq:balmer_dec}
    C(\hb) = \frac
    {\log_{10}{\left(\frac{\ha}{\hb}\right)}_{\rm theo} - 
     \log_{10}{\left(\frac{\ha}{\hb}\right)}_{\rm obs}
    }
    {0.4 \left[k(\lambda_{\ha}) - 
    k(\lambda_{\hb})\right]
    },
\end{equation}
where $\left(\frac{\ha}{\hb}\right)_{\rm obs}$ is the observed ratio between $\ha$ and $\hb$, and $\left(\frac{\ha}{\hb}\right)_{\rm theo}$ is the theoretically expected ratio of $2.86$ \citep[assuming case~B recombination, an electron density of $100$~cm$^{-3}$, and an electron temperature of $10^4$~K;][]{2006OsterbrockFerland}. This value is relatively insensitive to the assumed electron temperature, with variations of a factor of ${\sim}2$ causing changes of ${\sim}0.1$~mag in $V$-band extinction. The extinction coefficient at a given wavelenngth $\lambda$ is $k(\lambda) = A(\lambda)/{E(B-V)}$. We assume the standard Milky Way value for the total to selective extinction, $R_V = 3.1$. We use the extinction curve of \cite{1994ODonnell}, calculated using {\tt pyneb}\footnote{\url{https://pypi.org/project/PyNeb/}} \citep{2015Luridiana}. We also tested the THEMIS \citep{2014Kohler,2017Jones} extinction curve, which leads to ${<}0.01$~dex differences in the metallicities (using the same metallicity calibration). We therefore expect the choice of extinction curve will not qualitatively change our results. Values of $C(\hb) < 0$ typically have low signal-to-noise, and so we set them to 0. This has a negligible impact on our final maps, as these values typically fail our later signal-to-noise cuts. The corrected fluxes are then given by
\begin{equation}
    I_{\rm corr}(\lambda) = I_{\rm obs}(\lambda) \times 10^{0.4\,C(\hb) k(\lambda)}.
\end{equation}

Next, we make a number of cuts to ensure we select only star-forming regions, where metallicity prescriptions are appropriate. Firstly, for the strong lines used in this work ($\ha$, $\hb$, $\oiii$, $\nii$, and $\sii$), we require a signal-to-noise $\mathrm{(S/N)} > 5$. Secondly, we remove any regions with velocity dispersions $> 100~{\rm km\,s^{-1}}$ following \cite{2019Kreckel}, to ensure we remove supernova remnants. Thirdly, we use the following two Baldwin--Phillips--Terlevich \citep[BPT;][]{1981BaldwinPhillipsTerlevich} cuts to select star-forming regions. The first is the constraint from \cite{2003Kauffmann} for the $\nii$ diagram (left panel of Fig.~\ref{fig:ngc3627_bpt}):
\begin{equation}
    \log_{10}\left(\frac{\oiii}{\hb}\right) <  
    \frac{0.61}{\log_{10}\left(\frac{\nii}{\ha}\right) - 0.05} + 1.3.
\end{equation}
We also use the constraint from \cite{2001Kewley} for the $\sii$ diagram (right panel of Fig.~\ref{fig:ngc3627_bpt}):
\begin{equation}
    \log_{10}\left(\frac{\oiii}{\hb}\right) < 
    \frac{0.72}{\log_{10}\left(\frac{\sii}{\ha}\right) - 0.32} + 1.3.
\end{equation}
Both constraints must be satisfied for a pixel to be included in the final map. 

Because much of the DIG has line ratios consistent with photoionisation by massive stars \citep[$\sim$60\%;][]{2021Belfiore}, these cuts will not remove this diffuse component. We therefore use region masks from \cite{2021Santoro}, to isolate pixels dominated by \hii\ regions. This work gives a full account of how regions are defined, but briefly these regions are morphologically identified from MUSE H$\upalpha$ images, using {\sc HIIphot} \citep{2000Thilker}. Following this, integrated spectra within the regions are extracted and fitted using the DAP, and bona-fide \hii\ regions selected using similar criteria to ours described above (primarily velocity dispersion and BPT cuts). We only include these regions, and this has the effect of removing ${\sim}50$~per~cent of our pixels, mostly from just outside \hii\ regions (i.e.\ radiation leaking from the region). Based on visual inspection of the masks, the pixels removed in this step typically form rings around \hii\ regions, rather than small circular regions that would indicate our per-pixel cuts are detecting \hii\ regions that the morphological classification does not. Conversely, we also do not see pixels within the \hii\ masks that do not make our earlier selection criteria. This indicates a good level of agreement for the earlier selection criteria between per-pixel measurements and integrated spectra of these \hii\ regions. This masking step is applied to our convolved and regridded maps, simply by regridding to our larger pixel scale and world coordinate system. We will later use the pixel maps with all constraints except \hii\ region masking, to assess how including the DIG component affects our results (Sect.~\ref{sec:mcjack}). This will allow us a baseline comparison to future studies with potentially poorer spatial resolution, where removing the DIG component may not be possible.

\subsubsection{Metallicity Calibration}\label{sec:metallicity_calibration}

Our fiducial metallicity calibration is the \cite{2016PilyuginGrebel} S\nobreakdash-\hspace{0pt}calibration (hereafter Scal). This calibration uses three standard line diagnostics:
\begin{equation}
\begin{split}
    N_2 &= I_{\niijoin} / I_{\hb},\\
    S_2 &= I_{\sii} / I_{\hb},\\
    R_3 &= I_{\oiiijoin} / I_{\hb}.
\end{split}
\end{equation}
\noindent Note that for $\oiiijoin$ and $\niijoin$ we use only the flux of the stronger of the two lines, and we adopt a fixed theoretical ratio of ${3\!:\!1}$. The Scal prescription takes the form
\begin{multline}
    \met = a_1 + a_2\log_{10}(R_3/S_2) + a_3\log_{10}(N_2) \\
    + [a_4 + a_5\log_{10}(R_3/S_2) + a_6\log_{10}(N_2)] \times \log_{10}(S_2),
\end{multline}
where the coefficients $a_i$ are defined separately for the upper ($\log_{10}(N_2) \geq 0.6$) and lower ($\log_{10}(N_2) < 0.6$) branch. These are $a_i = [8.424, 0.030, 0.751, -0.349, 0.182, 0.508]$ for the upper branch and $a_i = [8.072, 0.789, 0.726, 1.069, -0.170, 0.022]$ for the lower branch. We use this as our fiducial metallicity calibration, as it shows small intrinsic scatter compared to direct metallicity measurements from auroral lines \citep{2016PilyuginGrebel, 2019aHo}. We repeat the analysis using the \cite{2016Dopita} calibration (see Sect.~\ref{sec:mcjack}), to assess the quantitative effect of using a different metallicity calibration. This calibration is given by
\begin{equation}
    y = \log_{10}\left(\frac{\nii}{\sii}\right) + 0.264\log_{10}\left(\frac{\nii}{\ha}\right),
\end{equation}
and then
\begin{equation}\label{eq:dopita_metallicity}
    \met = 8.77 + y + 0.45(y+0.3)^5.
\end{equation}
For each pixel that satisfies all of our cuts in Sect.~\ref{sec:data_prep}, we apply this metallicity calibration, to obtain a `pixel metallicity map'. We propagate through the uncertainties using the Python {\tt uncertainties} package, which rigorously accounts for errors in extinction correction and fluxes in the line diagnostics. We found some values to be anomalously low \citep[this is also seen in the catalogues of][]{2021Santoro}, and so remove outlier metallicities below $\met < 7.8$.

\begin{figure}
\includegraphics[width=\columnwidth]{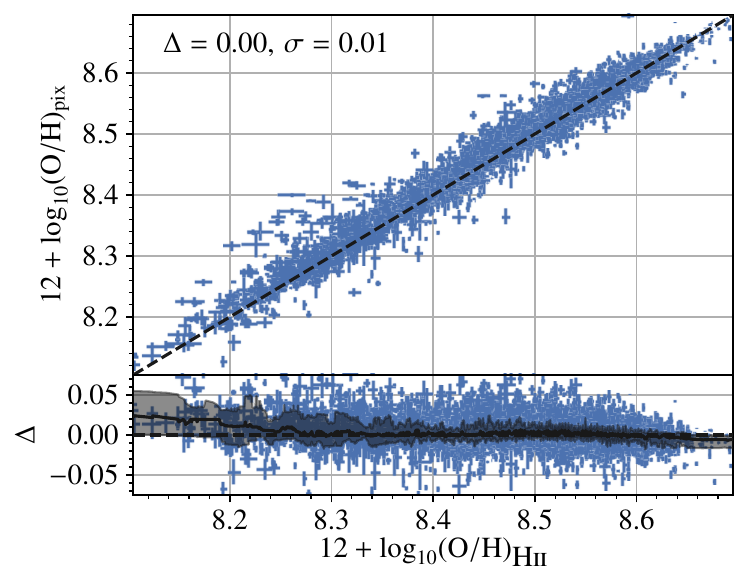}
\caption{$\ha$-flux weighted mean metallicities compared to integrated \hii\ region metallicites across our full galaxy sample. We see there is no systematic offset between these two measurements, and an extremely small scatter of $0.01$~dex. The lower panel shows the difference between each value, with the grey line indicating the rolling median and the shaded region the 16th and 84th percentiles of the distribution. There is a small deviation at the lowest metallicities which is slightly higher (and slightly lower at the highest metallicities), but these two regimes are dominated by low number statistics.} 
\label{fig:nebulae_pix_comparison}
\end{figure}

\begin{figure*}
\includegraphics[width=1.8\columnwidth]{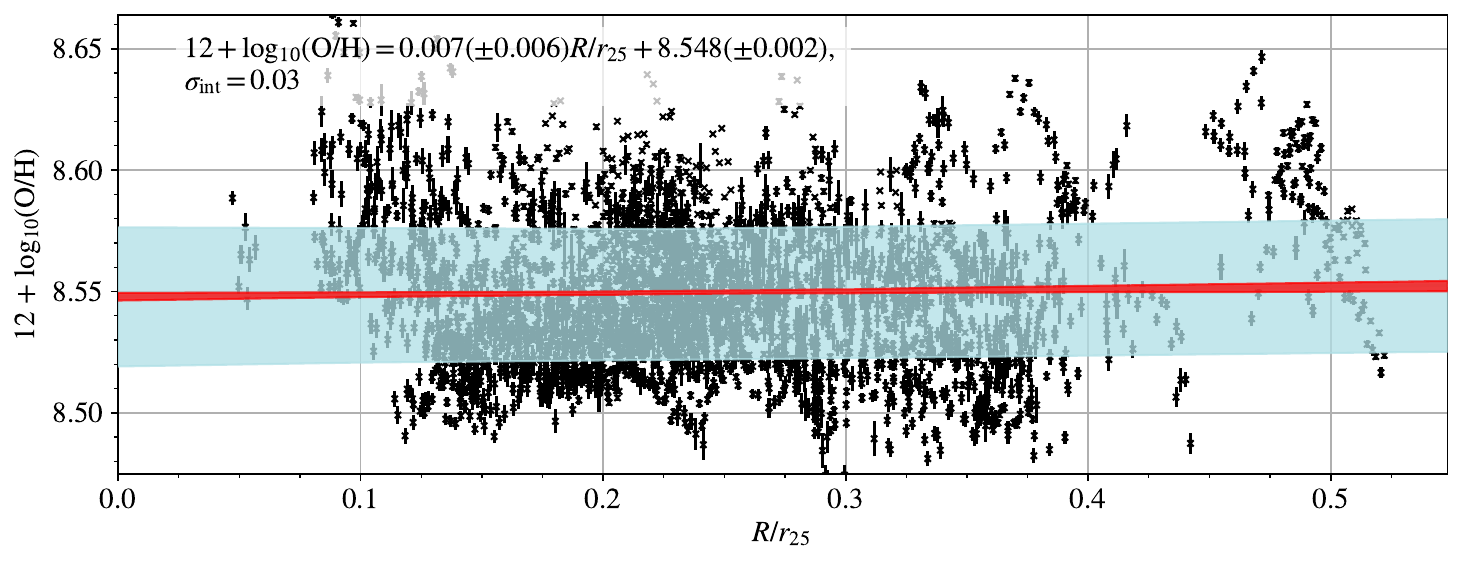}
\caption{Radial fit for pixel metallicities in NGC~3627. The parameters for this fit are given in the top left. The shaded blue region highlights the intrinsic scatter. The shaded red region shows the uncertainty in the regression, but many points lie significantly beyond this region and have clear structure, highlighting the presence of higher-order variations with respect to the one-dimensional radial trend.} 
\label{fig:radial_fit}
\end{figure*}

As an additional check of the robustness of our metallicity measurements, we test against the metallicities obtained for integrated \hii\ regions, using the catalogue from \cite{2021Santoro} and the same metallicity calibration. We reproject each of these masks onto our convolved, regridded maps, and for each region as defined in this catalogue, we take the $\ha$-flux weighted mean metallicity of all our pixels falling within the spatial extent of the \hii\ region. We find a mode of one pixel per \hii\ region, and a median of three, indicating that our choice to convolve to 120~pc resolution does indeed lead to, on average, a single \hii\ region per pixel. This comparison is shown in Fig.~\ref{fig:nebulae_pix_comparison}. We see an excellent agreement between these two approaches, with no systematic offset and a small scatter of $0.01$~dex, so we are confident in this `per-pixel' metallicity method going forwards.

\section{Two-Dimensional Metallicity Maps}\label{sec:gpr}

\begin{figure*}
\includegraphics[width=2\columnwidth]{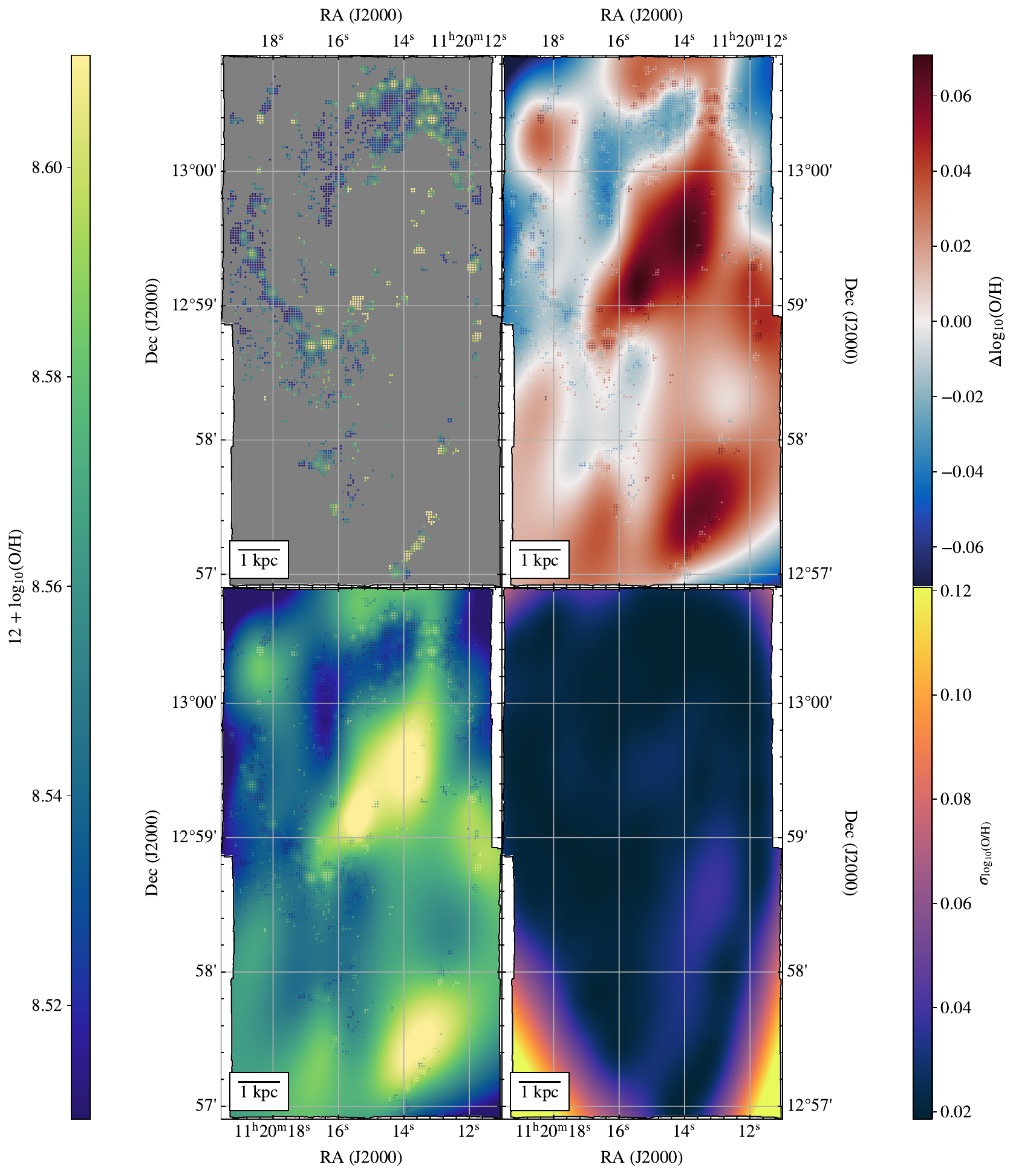}
\caption{{\it Top left:} Pixel metallicities for NGC~3627. This galaxy has a metallicity filling factor (Sect \ref{sec:mcjack}) of 14~per~cent, meaning its \hii\ region density is slightly higher than most of our galaxy sample. {\it Top right:} Residual metallicities after radial gradient is subtracted, overlaid on the GPR model for these points. {\it Bottom left:} Final 2D metallicity map, showing clear higher-order variation. {\it Bottom right}: Associated metallicity error map, combining both uncertainties in the radial fit, as well as the higher-order GPR.} 
\label{fig:gpr_fit}
\end{figure*}

\begin{figure*}
\includegraphics[width=1.8\columnwidth]{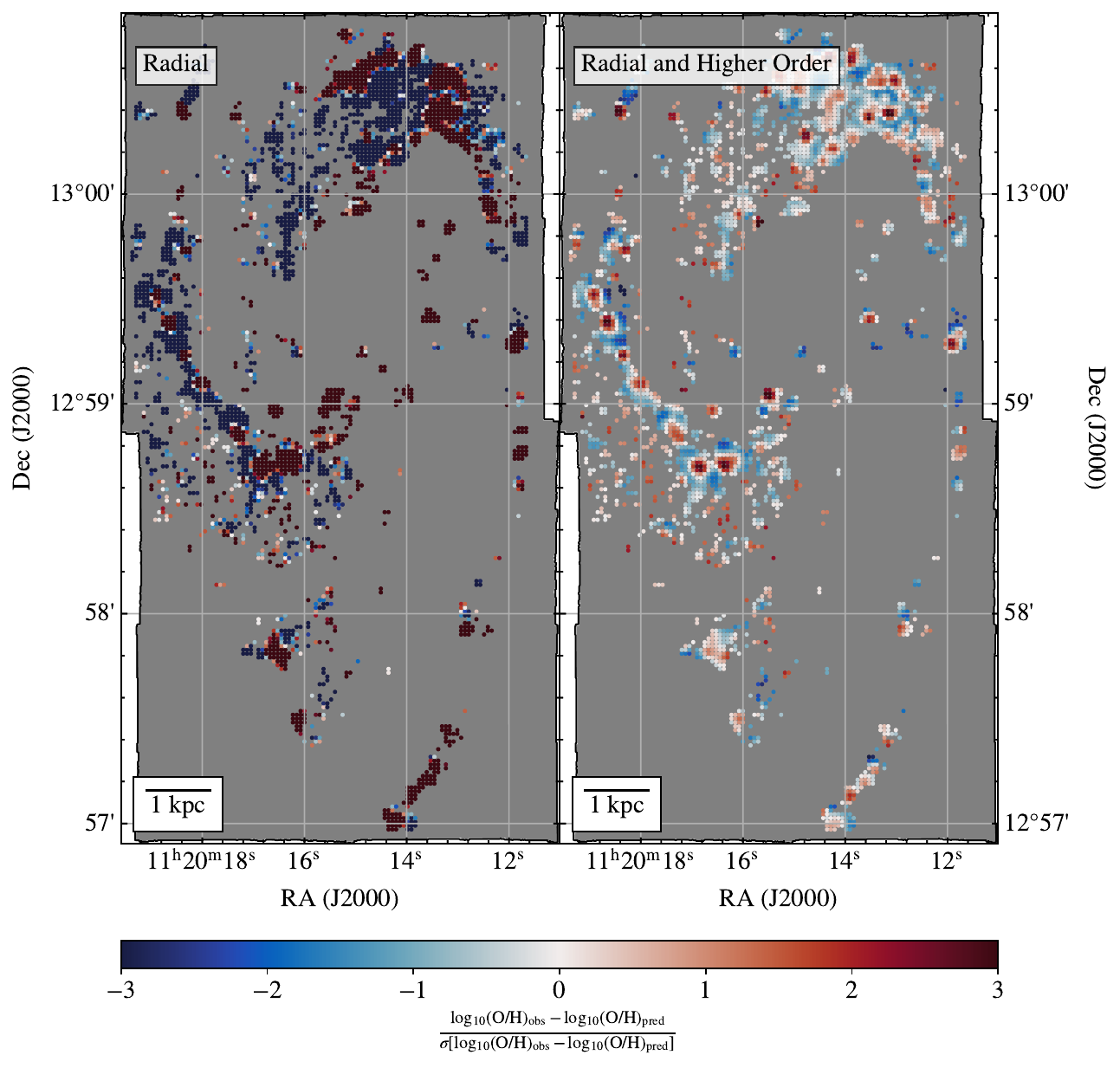}
\caption{Map of the residual metallicity (i.e.\ $\met_\mathrm{obs} - \met_\mathrm{pred}$) normalised by the combined errors in the observations and model fits of NGC~3627 for {\it left}: a radial gradient and {\it right}: our full 2D model. White indicates where the model and observations agree -- including the higher-order variations produces a better fit to the data, reducing many highly significant outliers to less than $3\sigma$.}
\label{fig:residual_map}
\end{figure*}

\begin{figure*}
\includegraphics[width=1.9\columnwidth]{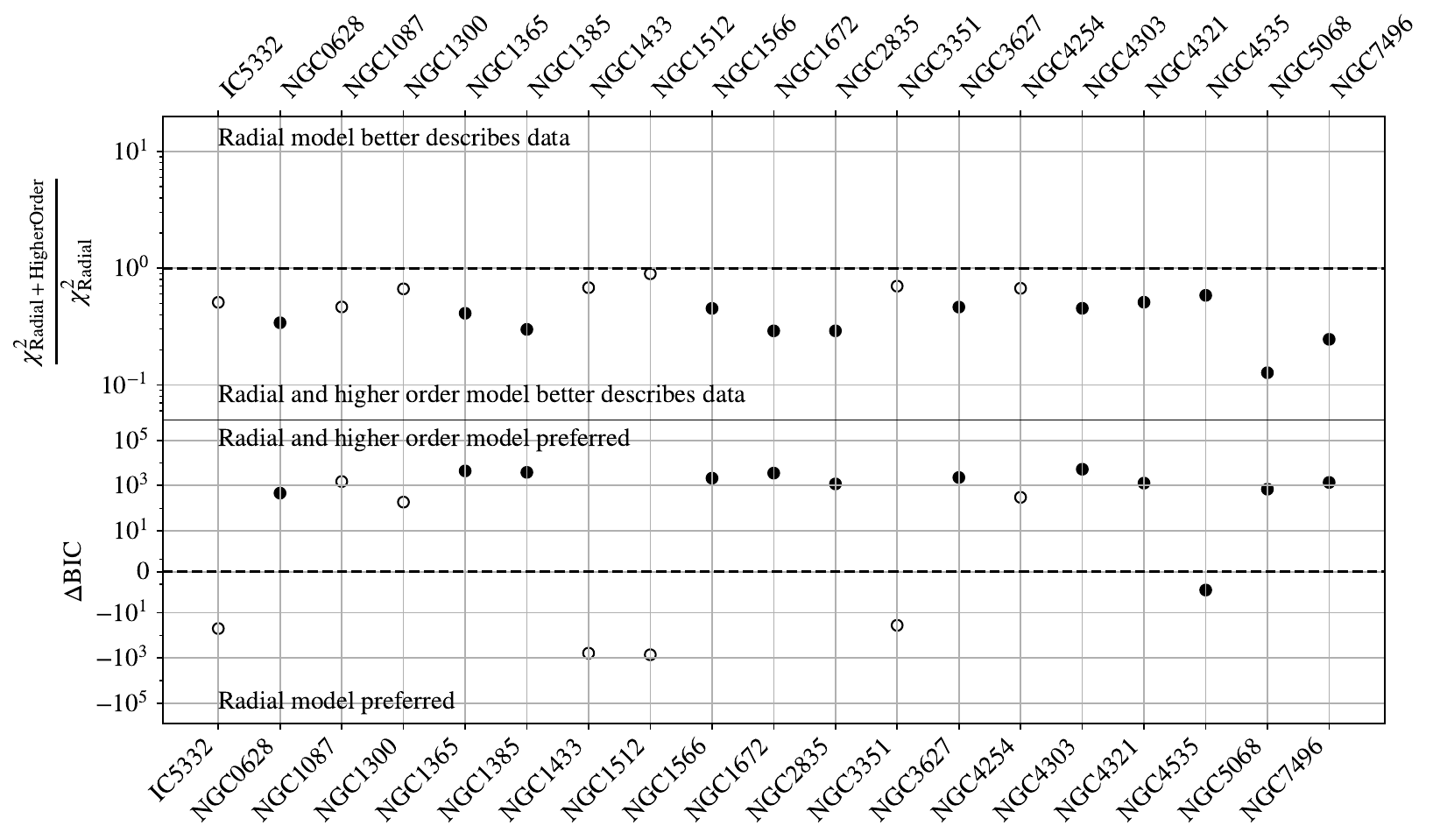}
\caption{Figures of merit for the fitting procedures. {\it Top}: Ratio of $\chi^2$ values for the GPR (radial+higher-order) and radial-only model. Above one, the radial-only fit is preferred, and below one, the combined fit preferred. {\it Bottom}: Difference in BIC~statistics for the two models. Above zero, the combined model is preferred, below zero, the radial-only model is preferred. Points are filled if they pass our later significance testing (Sect.~\ref{sec:mcjack}), and unfilled if they do not. The dashed black line indicates where the models are equally preferred, in each case.} 
\label{fig:model_comparisons}
\end{figure*}

\subsection{Fitting the pixel metallicities}

Using the pixel metallicity maps obtained in Sect.~\ref{sec:metallicity_calibration}, we use a simple model to describe the underlying galactic metallicity distribution. We first assume the bulk of the variation is driven by a radial dependence, which has been typically employed in the literature \citep[e.g.][]{1971Searle, 2015Ho}. Throughout this, we will fit in the coordinate frame of the galaxy, by calculating a deprojected galactocentric radius to each pixel (using the inclination and position angles listed in Table \ref{tab:galaxy_params}). We note that this will lead to a slightly noncircular beam in the deprojected frame, but as our galaxies are relatively face-on this will be negligible. We fit a simple model taking into account the measured uncertainties on each metallicity, as well as an intrinsic scatter which captures the scatter between points that is not explained by the statistical uncertainties in the metallicity measurements, and is typically on the order of 0.03~dex. The \mbox{(ln-)}\linebreak[0]{}likelihood function of this model is \citep[following][]{2010Hogg}
\begin{equation}
    \ln(\mathcal{L}) = -\frac{1}{2} \sum_n \left[ \frac{(Z_n - m \times R/r_{25, n} - Z_0)^2}{s_n^2} + \ln\left(2 \pi s_n^2\right)\right],
\end{equation}
where $Z_n$ and $R/r_{25, n}$ are the metallicity and galactocentric radius for each pixel, respectively, $m$ is the radial metallicity gradient, $Z_0$ the metallicity at $R=0$, and $s_n^2 = \sigma_{n,\,\mathrm{meas}}^2 + \sigma_\mathrm{int}^2$, the quadratic sum of the measured error and intrinsic scatter of the metallicity. We maximise this likelihood function using \texttt{emcee} \citep{2013emcee}, with 500~steps and 500~walkers. We remove the first half of these steps as `burn-in', and calculate our best-fitting parameters from the remaining samples. An example radial fit is shown in Fig.~\ref{fig:radial_fit}, for NGC~3627. Clearly, many measurements lie significantly off this simple radial model (see also Fig.~\ref{fig:residual_map}), motivating the need for higher-order fitting terms.

To model this two-dimensional variation, we use a Gaussian Process Regression (GPR) technique \citep[for a mathematical introduction to this process, we refer the reader to][]{2006RasmussenWilliams}. GPR is a probabilistic interpolation that models the covariance between neighbouring points using a covariance kernel, and thus is well-suited to the task of modelling the metallicity distribution of galaxies, as \hii\ regions have highly correlated metallicities over relatively small spatial scales \citep[on the order of 100~pc;][]{2020Kreckel}. GPR has been shown to recover the underlying distributions in a minimally biased manner \citep{2019GonzalezGaitan}. As GPR is a Bayesian method, it produces a posterior probability distribution function at each position, allowing us to calculate an uncertainty for each interpolated value. This technique has previously been used to produce metallicity maps by \cite{2019Clark}, who found the GPR to reliably recover metallicity values (see their appendix~C). GPR is a commonly applied machine learning technique, and has recently seen an increase in the number of applications in other astronomical contexts, particularly to model the lightcurves of transiting exoplanets \citep[e.g.][]{2017Prsa, 2019Espinoza}, as well as 3D modelling of dust in the Milky Way \citep{2019Green}.

Our covariance kernel is the Mat{\'e}rn kernel, which is a standard choice for GPR modelling of 2D data \citep{2006RasmussenWilliams}. The choice of kernel will have an impact on the final map \citep[see especially figs.~4.1 to~4.4 in][]{2006RasmussenWilliams}, but our choice of kernel here allows for variation from quite granular to smooth maps, in an attempt to cover the potential metal distributions within these galaxies. The Mat{\'e}rn kernel appears similar to a Gaussian, but with broader tails (sensitive to covariance over larger scales), and a narrower peak (sensitive to covariance on short distances). It takes the form
\begin{equation}\label{eq:matern}
    k(r_i, r_j) = 
    \frac{1}{\Gamma(\nu) 2^{\nu -1}}
    \left(\frac{\sqrt{2\nu}}{\sigma_{l, \rm obs}}d(r_i, r_j)\right)^\nu
    {\rm K}_\nu \left(\frac{\sqrt{2\nu}}{\sigma_{l, \rm obs}}d(r_i, r_j)\right),
\end{equation}
where $r_i, r_j$ are the two-points being considered, $d$ the distance between them, ${\rm K}_\nu$ a modified Bessel function, and $\Gamma(\nu)$ the gamma function. The kernel therefore only has two parameters, $\nu$, which determines the smoothness of the kernel, and $\sigma_{l, \rm obs}$, the kernel length scale. This kernel length scale is related to the metal mixing scale (although, we will later show the absolute values differ by a significant factor). We fix $\nu$ to a value of~$1.5$, which makes the computation significantly more efficient, as equation~\eqref{eq:matern} then simplifies to
\begin{equation}\label{eq:matern_simplified}
    k(r_i, r_j) = 
    \left(1 + \frac{\sqrt{3}d(r_i, r_j)}{\sigma_{l, \rm obs}}\right)
    \exp\left(-\frac{\sqrt{3}d(r_i, r_j)}{\sigma_{l, \rm obs}}\right).
\end{equation}
This approach is practically effective \citep{2006RasmussenWilliams}, and has been used to create two-dimensional metallicity maps in previous work \citep{2019Clark}. This choice of $\nu$ imposes a level of smoothness in the final maps, but we find that experimenting with $\nu=0.5$ leads to an extremely noisy map, and the kernel length scale is not constrained \citep[indeed, this is noted in sect.~4.2 of][]{2006RasmussenWilliams}. We therefore proceed using the standard $\nu=1.5$, but note that this choice of smoothness parameter may lead us to miss some variation on very small spatial scales (see especially the structure in the right panel of Fig.~\ref{fig:residual_map}).

We perform the fitting to the radially subtracted metallicity maps, using the {\tt Gaussian\-Process\-Regressor} in {\tt scikit-learn}, a {\sc python} package for machine learning. We set prior bounds of $[0.001, 5]\,r_{25}$ on~$\sigma_{l, \rm obs}$ (where $r_{25}$ is the 25$^{\rm th}$ mag isophotal contour, a measure of the galaxy size; see Tab \ref{tab:galaxy_params}), to allow freedom from extremely granular higher-order metallicity maps (this range includes the resolution of the maps) to extremely smooth distributions. The MUSE field of view (FOV) typically extends to around $1r_{25}$, and so a kernel length scale larger than this may not be recoverable. We performed some simple one-dimensional tests sampling from a Gaussian process with a number of kernel length scales (from 0 up to 10 times the FOV of the data), adding some random noise at the 1\% level (similar to our data). We found the kernel length scale was accurately recovered up to around five times the data FOV, and above this tended to underestimate the true value, motivating our choice of upper prior bound limit. For a kernel length scale hitting the upper prior bound, this indicates the GPR finds no significant covariance between points in the map, and not that the kernel length scale has been constrained. As the MUSE maps do not cover the entirety of the galaxy, they will not be sensitive to the typically higher scatter at much higher galactocentric radii \citep[e.g.][]{2014Pastorello}. As such, this kernel scale length should be interpreted as an average of any variations in the mixing scaling within our FOV, and may not be representative of a mixing scale in the outskirts of galaxies. Our final metallicity map is the linear combination of the radial fit and GPR fit. This means that in regions with few measurements, the metallicity map will tend to the radial gradient, rather than $\met=0$. Our uncertainties combine the individual uncertainties from the radial and GPR fit added in quadrature. An example is shown in Fig.~\ref{fig:gpr_fit}, for NGC~3627, and in Appendix \ref{app:all_fit_overviews} we show the equivalent plot for all 19 galaxies in the PHANGS-MUSE sample. Fig.~\ref{fig:residual_map} shows the improvement by including the higher-order terms in this fitting, also for NGC~3627.

We use two figures of merit to judge whether including this higher-order term improves the quality of the fit to the observed metallicities. The first is the standard $\chi^2$~metric, to quantify how well the model describes the data, and the second is the Bayesian Information Criterion (BIC), which accounts for the number of parameters in the models, and penalises more complex models, even if they do describe the data better. We show (a) the ratio of $\chi^2$ values and (b) the difference in BIC ($\Delta{\rm BIC}$) statistics in Fig.~\ref{fig:model_comparisons}, to show which model is preferred by each statistic. In all cases, including the two-dimensional information leads to a fit where the $\chi^2$ ratio prefers including the GPR fit. This is also the case for the majority of galaxies when using $\Delta{\rm BIC}$. The BIC statistic strongly favours including the GPR fitting for 14 of the galaxies (generally, a $\Delta{\rm BIC}>10$ is considered strong evidence to prefer one model over another). However, there are five galaxies where the radial-only model is preferred. Of these, four strongly prefer the radial-only model over including the GPR fit, and NGC~4535 is a marginal case. In Sect.~\ref{sec:mcjack}, we will describe how we determine if the higher-order component is deemed significant.

\subsection{Determining the significance of higher-order variations}\label{sec:mcjack}

\begin{figure}
\includegraphics[width=\columnwidth]{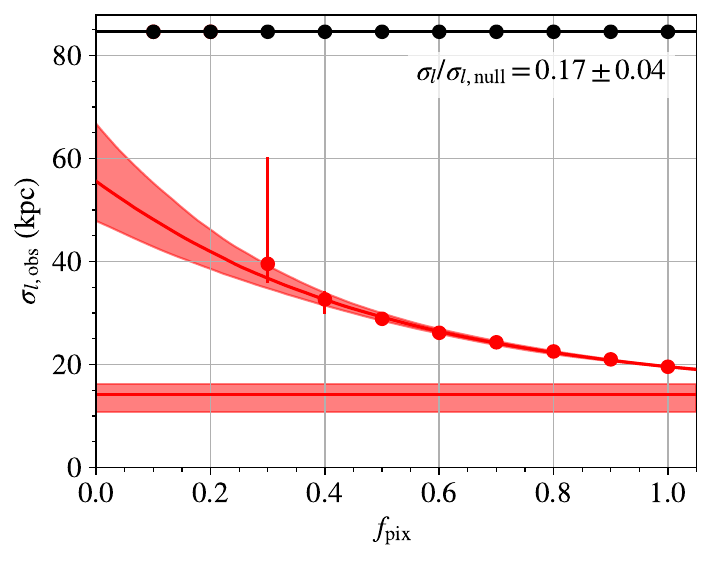}
\caption{Monte Carlo Jackknife (McJack) method (Sect.~\ref{sec:mcjack}) to correct the measured metallicity kernel length scale in NGC~3627. The horizontal red line shows the extrapolation of the line fitted to the real data (red points) out to infinity (our corrected kernel length scale, $\sigma_l$), the black one to our null hypothesis ($\sigma_{l, \rm null}$, which is always $5r_{25}$). Clearly, the data show a significantly different kernel length scale to the null hypothesis (highlighted by the text in the top right), and so we deem this higher-order variation to be significant.} 
\label{fig:scale_length_corr}
\end{figure}

\begin{figure}
\includegraphics[width=\columnwidth]{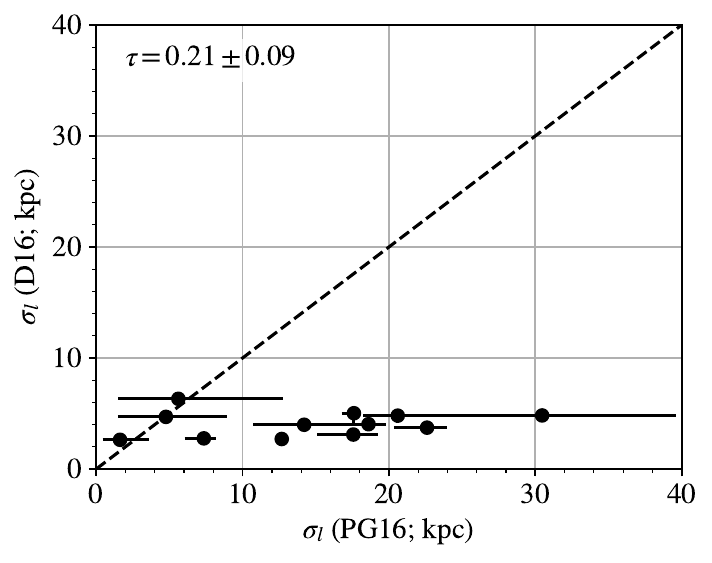}
\caption{Comparison between kernel length scales derived using the \citet{2016Dopita} or the \citet{2016PilyuginGrebel} Scal metallicity calibration. The ${1\!:\!1}$ relationship is shown as a dashed black line. Kendall's $\tau$ correlation coefficient is shown in the top left. Typically, the kernel length scales derived from the \citet{2016Dopita} are significantly shorter, and reflect the higher scatter in this metallicity calibration, which the GPR models as significant variation over short scales.}
\label{fig:pg16_d16_scale_length_comparison}
\end{figure}

\begin{figure*}
\includegraphics[width=2\columnwidth]{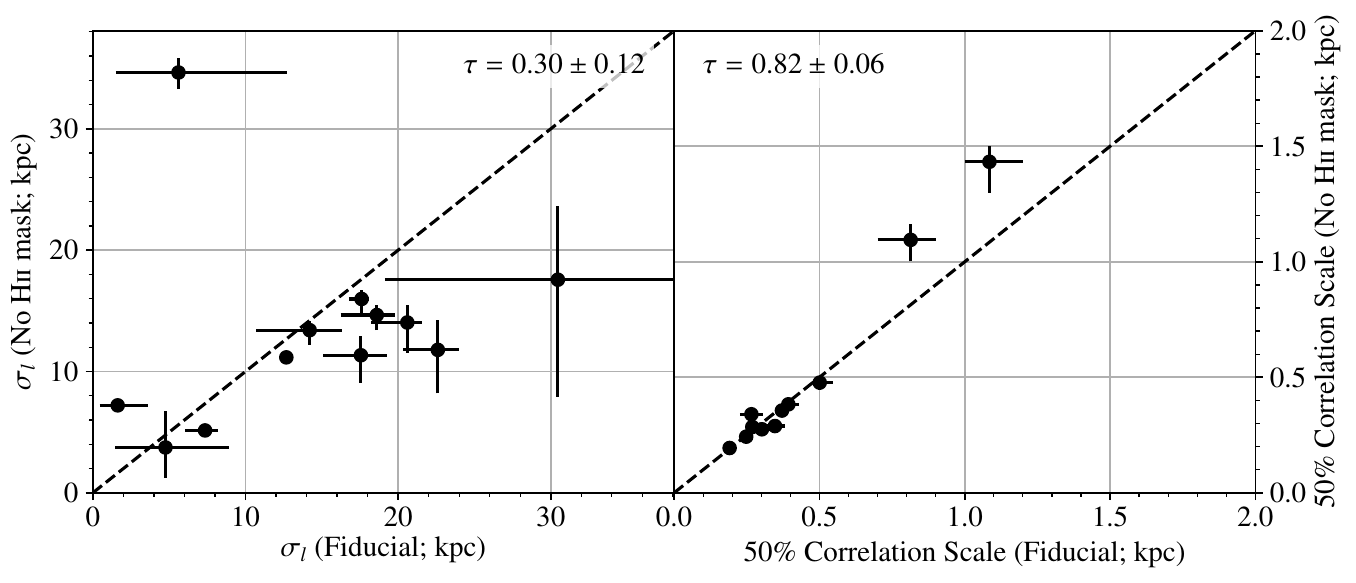}
\caption{{\it Left:} kernel length scales calculated from our fiducial assumptions on which pixels to include, compared to when we relax our restriction to only include pixels within \hii\ regions as defined by \citet{2021Santoro}. {\it Right:} the same exercise but for the 50\% correlation scales of the two-point correlation function. In each case, the ${1\!:\!1}$ line is shown as a dashed black line, and we show Kendall's $\tau$ correlation coefficient.} 
\label{fig:fiducial_no_hii_mask_comparison}
\end{figure*}

\begin{figure}
\includegraphics[width=\columnwidth]{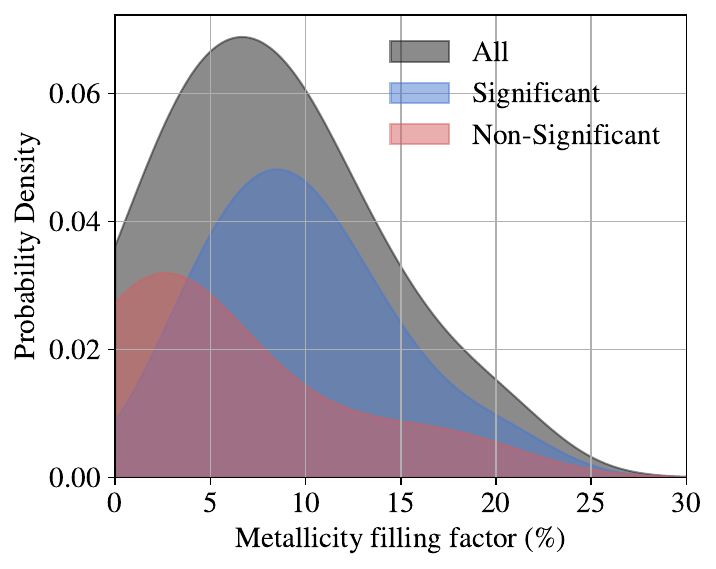}
\caption{Distribution of metallicity filling factors for galaxies in which we measure a significant higher-order variation (blue) and galaxies where we do not (red). The overall distribution for all galaxies is shown in grey in the background. Due to the relatively small number of galaxies, we show the distribution as a Kernel Density Estimate (KDE), to aid visualisation. We typically see a lower filling factor for galaxies where we do not measure significant higher-order variations.} 
\label{fig:filling_factor}
\end{figure}

\renewcommand{\arraystretch}{1.25}
\input{scale_lengths_pg16_scal}
\renewcommand{\arraystretch}{1}

\begin{figure*}
\includegraphics[width=1.8\columnwidth]{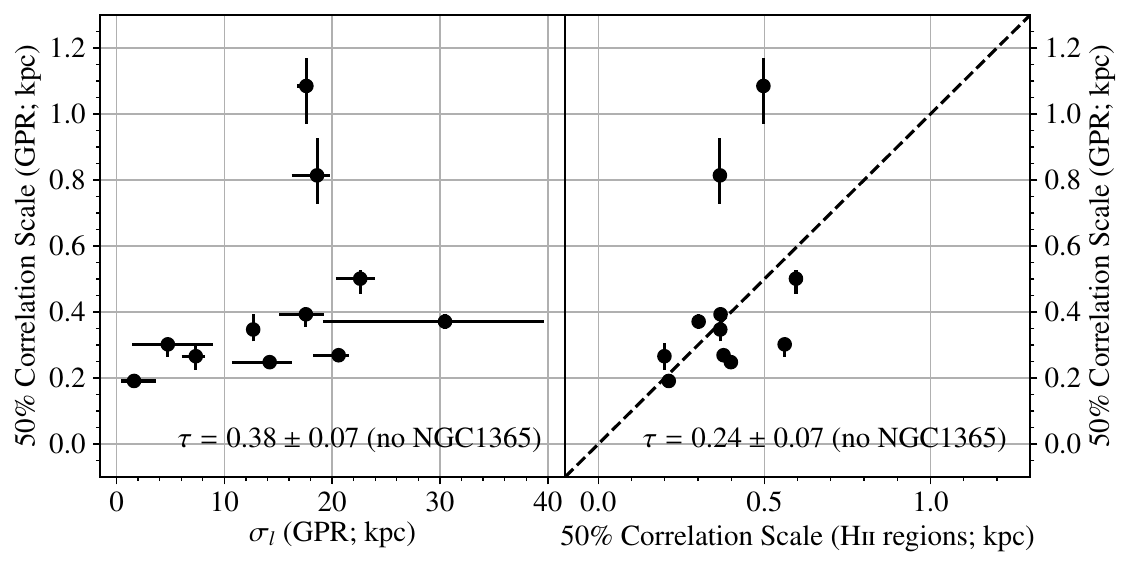}
\caption{{\it Left}: Comparison between derived kernel length scales and 50~per~cent correlation scale from two-point correlation function. The 50~per~cent correlation scales are significantly shorter than the kernel length scales, by an order of magnitude or more. {\it Right}: Comparison between 50~per~cent two-point correlation functions derived in this work, and for the same galaxies from the nebulae catalogues of \citet[][Appendix~\ref{app:two_point_hii}]{2021Santoro}. The dashed black line indicates the ${1\!:\!1}$ relation. In each case, we show Kendall's $\tau$ correlation coefficient excluding NGC~1365 (the notable outlier).}
\label{fig:two_point_corr}
\end{figure*}

As we do not have a metallicity measurement for every pixel in our map, the covariance kernel length scale may be somewhat overestimated (with fewer values, the GPR is unable to reliably model the covariance between points, and so tends towards larger kernel length scales, indicating smoother maps). To correct for this bias, we perform a nested Monte Carlo Jackknife approach (`McJack' for short). We retain a certain fraction of pixels ($f_{\rm pix}$ from our original pixel metallicity maps, from $0$ to $100$~per~cent, where $100$~per~cent corresponds to $N_Z$ in Table \ref{tab:galaxy_params}) in steps of $10$~per~cent and refit the GPR. In this sense, we can extrapolate beyond $f_{\rm pix}=1$, as this is set by the maximum number of metallicity measurements in our pixel map. This is similar to cross-validation in machine learning models, with a subset of the data being held back to verify the robustness of the output parameters. For each fraction, we also perturb the metallicities of the remaining pixels by their associated error, to estimate an uncertainty on the measured kernel length scale. For each fraction, we perform these perturbations a total of $100$~times. The change in $\sigma_l$ with $f_{\rm pix}$ appears to be well-described by a negative exponential, of the form
\begin{equation}
    \sigma_{l, \rm obs} = A\exp({-\lambda f_{\rm pix}}) + \sigma_{l},
\end{equation}
where $A$ is the amplitude of the function at $f_{\rm pix}=0$, $\lambda$ is an exponential scale length, and $\sigma_{l}$ the corrected kernel length scale (the extrapolation of this model to $N_{\rm all} \gg N_Z (f_{\rm pix} \to \infty)$, an estimate of the case where we have a metallicity for every pixel in the map). The model choice here is motivated by the shape of the data, rather than physically; we experimented with a number of declining functional forms (e.g.\ half-normal, half-student~T), and found that the exponential best describes the trends seen in all cases. We could instead use $f_{\rm pix}$ to refer to the total number of pixels in the map ($N_{\rm pix}$ in Table \ref{tab:galaxy_params}); this would change $\lambda$ but keep $\sigma_{l}$ identical. We note that we only fit points that are not at (or consistent with) the upper prior bound of $5r_{25}$, and where we have more than 3~points to fit (so we have at least one degree of freedom). An example of this fit is shown in Fig.~\ref{fig:scale_length_corr}, for NGC~3627.

Having corrected our kernel length scale, we next turn to the question of determining whether the determined kernel length scale is statistically significant with respect to a map with no correlation between neighbouring pixels. For this, we perform a null hypothesis test which is essentially identical to the McJack described in the previous paragraph, but before any pixels are removed we randomly shuffle them around the map after radial gradient subtraction. In this case, the GPR always hits the upper bound of the prior, indicating a smooth map with no real structure. If $\sigma_{l}$ is significantly different to this null hypothesis (above the $1\sigma$ level), then we conclude we have measured statistically significant higher-order variations. Of our 19 galaxies, we find that \ngal\ have significant higher-order variations, and that typically those that do not have significant higher-order variations are those that do not have large changes in their $\chi^2$ values (see Fig.~\ref{fig:model_comparisons}).

We make two checks on these kernel length scales to see how some of our assumptions may affect them. In the first, we repeat this entire procedure using the \cite{2016Dopita} metallicity calibration, see Equation~\eqref{eq:dopita_metallicity}, and we show this in Fig.~\ref{fig:pg16_d16_scale_length_comparison} for galaxies where the $\sigma_l$ values measured from both metallicity calibrations are deemed to be significant. The correlation between the kernel length scales derived with these two different calibrations has Kendall's $\tau$ correlation coefficient of $0.21\pm0.09$. This is similar to the correlations seen by \cite{2021Li} between scale lengths measured from different metallicity calibrations. \cite{2021Li} use the Pearson correlation coefficient, but repeating this exercise with that correlation coefficient yields nearly identical results. However, the kernel length scales measured are typically quite different, with the Scal values being significantly higher. This reflects the increased scatter of metallicities between neighbouring \hii\ regions in the \cite{2016Dopita} calibration, which the GPR models as significant variation over small scales.  We adopt Scal as our fiducial metallicity calibration due to its low scatter with respect to direct metallicity measurements in an attempt to avoid this issue, but we note that maps derived from different calibrations will produce (sometimes significantly) different looking maps \cite[see also appendix~C of][]{2019Kreckel}.

For our second test, we relax our constraint on limiting to pixels within \hii\ regions as defined from their $\ha$ morphology by \cite{2021Santoro}. We repeat our analysis, and show the derived kernel length scales from this exercise compared to our fiducial assumptions in Fig.~\ref{fig:fiducial_no_hii_mask_comparison}. There is a reasonable agreement, with Kendall's $\tau$ of $0.30\pm0.12$, indicating that including pixels that are dominated by DIG emission does not bias our results in a sample-wide sense, but including DIG emission can lead to significant differences in individual kernel length scales. We therefore urge caution in interpreting length scales measured on data where morphological classification of \hii\ regions is not possible.

The lack of significant higher-order variations in seven galaxies may be due to two reasons. The first is that these galaxies truly only possess a radial metallicity gradient, or second that our data are insufficient in number for the fitting algorithm to measure this variation. To test this, we calculate a `metallicity filling factor', which is simply the fractional area on the sky occupied by \hii\ regions. We show the distribution of these filling factors in Fig.~\ref{fig:filling_factor}. There may be local variations of the filling factor in radius and azimuth, which will manifest as larger uncertainties in the GPR fit. Clearly, for galaxies where we do not deem the higher-order component to be significant, we have far fewer points to fit the GPR to. This lack of higher-order structure is likely driven by the lack of a sufficient number of \hii\ regions within these galaxies, and not necessarily by a real lack of higher-order variations in the galaxy itself. Since these galaxies are the ones in which the GPR does not provide a significant improvement over simply a radial metallicity gradient, we will not use these in our later analysis. Our corrected kernel length scales are given in Table~\ref{tab:scale_lengths_corrected}. Our kernel length scales vary from $1.5$ to $30$~kpc, with a median value of $15$~kpc ($0.1{-}1.9r_{25}$, median $1.1r_{25}$). We find that there is a slight trend with the MUSE FOV size (Kendall's $\tau$ of $0.15\pm0.12$), but no trend with the galaxy inclination (Kendall's $\tau$ of $0.12\pm0.12$).

\subsection{Two-point correlation function}\label{sec:two_point_corr}

\cite{2020Kreckel} performed an analysis of the two-point correlation function of metals for a subset of eight galaxies of the PHANGS--MUSE sample, applied to measurements of individual \hii\ regions. They found a high correlation in metallicity over small scales, with the exception of IC~5332 (a~galaxy where we also do not detect significant higher-order variations). They quantify the mixing scale via a percentage correlation scale in the two-point correlation function (their sect.\,4). We apply the same measurement to our maps here, to see how comparable our results are to those applied only to \hii\ regions. The two-point correlation function at a spatial scale $r$ is given as
\begin{equation}\label{eq:two_point_corr}
    \xi (r) = \left\langle \frac{\overline{Z({\bf r_1})Z({\bf r_2})} - \overline{Z}^2}{\overline{\big(Z - \overline{Z}\big)^2}}\right\rangle,
\end{equation}
where $Z$ represents a metallicity measurement, and $|\mathbf{r_1}-\mathbf{r_2}| \le r$.  Horizontal lines indicate averaging over all \hii\ regions in the galaxy, while angle brackets indicate averaging over all choices of $\mathbf{r_1}$.  At $r=0$, each \hii\ region correlates perfectly with itself and a $100$\% correlation is recovered. We apply this to our radially-subtracted, GPR fitted maps, taking a small percentage (1~per~cent, leaving us a minimum of ${\sim}3000$ pixels to calculate this statistic) of the total pixels in our maps each time, to reduce computation time and allow us to determine jackknife errors. Of the remaining pixels, we also perturb these by the measured errors. We calculate the two-point correlation for a number of scales (from 0 to 5~kpc), and calculate the 50~per~cent correlation scale. We show the relation of the kernel length scales and 50~per~cent correlation scale in the left panel of Fig.~\ref{fig:two_point_corr}, and list these values in Table~\ref{tab:scale_lengths_corrected}. The kernel length scale and the 50~per~cent correlation scale are positively correlated with each other, but the kernel length scale is often more than an order of magnitude larger in absolute value. The difference in these two values reflects the fact that the 50~per~cent correlation function is a measure of small-scale, highly local correlation between values, whilst the kernel scale length is sensitive more to large-scale, smoother variations between more separated measurements. The two quantities are not monotonically related, and there is no simple transformation between them. NGC~1365 is an outlier in this comparison, with a 50~per~cent correlation scale of over 4~kpc. We do not include this in our plot, or calculation of the Kendall's $\tau$ coefficient, which is $\tau=0.38\pm0.07$. We also compare how the 50\% correlation scales changes based on whether we use \hii\ region masks to define pixels to fit or not. This is shown in the right panel of Fig. \ref{fig:fiducial_no_hii_mask_comparison}, and there is an excellent agreement between the values calculated using these two different assumptions, with a $\tau=0.82\pm0.06$ and most points clustered along the ${1\!:\!1}$ line. Whilst we have earlier motivated our choice to remove pixels outside \hii\ regions before fitting the GPR, this result highlights this choice will not bias our results significantly.

We also apply the two-point correlation function to the full PHANGS nebulae catalogues \citep[][i.e.\ only using \hii\ regions]{2021Santoro} in Appendix~\ref{app:two_point_hii} for all 19 galaxies in the PHANGS--MUSE sample, and we compare the values calculated from our interpolated radially-subtracted GPR maps to those calculated from the \hii\ region catalogues in the right panel of Fig.~\ref{fig:two_point_corr}, for those where we deem the measured kernel length scales to be significant, and excluding NGC~1365, as this also has a very high 50~per~cent correlation scale measured from the \hii\ regions. This likely indicates the scale length is primarily driven by \hii\ regions along the bar of this galaxy, which fills most of the MUSE FOV, rather than some inherent limitation in either the GPR fitting or applying a two-point correlation statistic to \hii\ regions. We see a good agreement between the 50~per~cent two-point correlation scales between the GPR map and from the \hii\ region catalogues, with many following the ${1\!:\!1}$ relationship, and determine a Kendall's correlation coefficient $\tau=0.24\pm0.07$. We are therefore confident that our maps reflect a similar covariance between neighbouring regions as those found when considering morphologically selected \hii\ regions. Furthermore, since the 50~per~cent correlation scales are similiar between these two methods, we are confident these different analyses characterise the metal mixing in a similar way. We will therefore use the 50~per~cent correlation scale from the two-point correlation function in our analysis going forwards, as this has been previously employed in the literature \citep{2020Kreckel}, and shown to produce similar correlation scales to the model of \cite{2018KrumholzTing}, as shown by \cite{2021Li}. However, as we have shown with our comparison between the kernel scale length and 50~per~cent correlation scale, using different measures of `scale lengths' can result is significantly different absolute values. In Sect.~\ref{sec:global_correlations} we will search for trends between this correlation scale and various global parameters.

\section{Environmental dependence of metal enrichment}\label{sec:environment}

\begin{figure}
\includegraphics[width=\columnwidth]{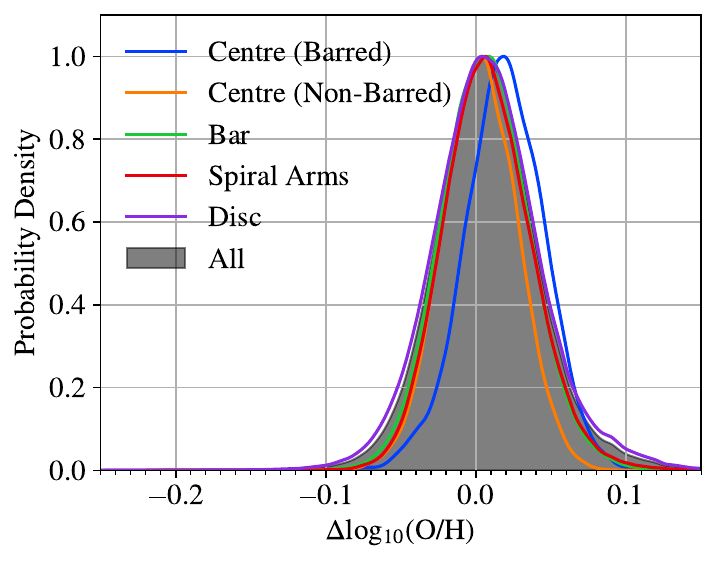}
\caption{$\Delta\oh_{\ha}$ for each environment for all \ngal\ galaxies with significant higher-order variations. The overall distribution is shown as shaded grey region, and the different environments with differently coloured lines. Each distribution is normalised to have a peak of~1. We separate the centres of barred and non-barred galaxies, as we may expect a different trend in abundance variation between barred and unbarred galaxies.}
\label{fig:abundance_offsets_kde}
\end{figure}

\begin{figure*}
\includegraphics[width=1.9\columnwidth]{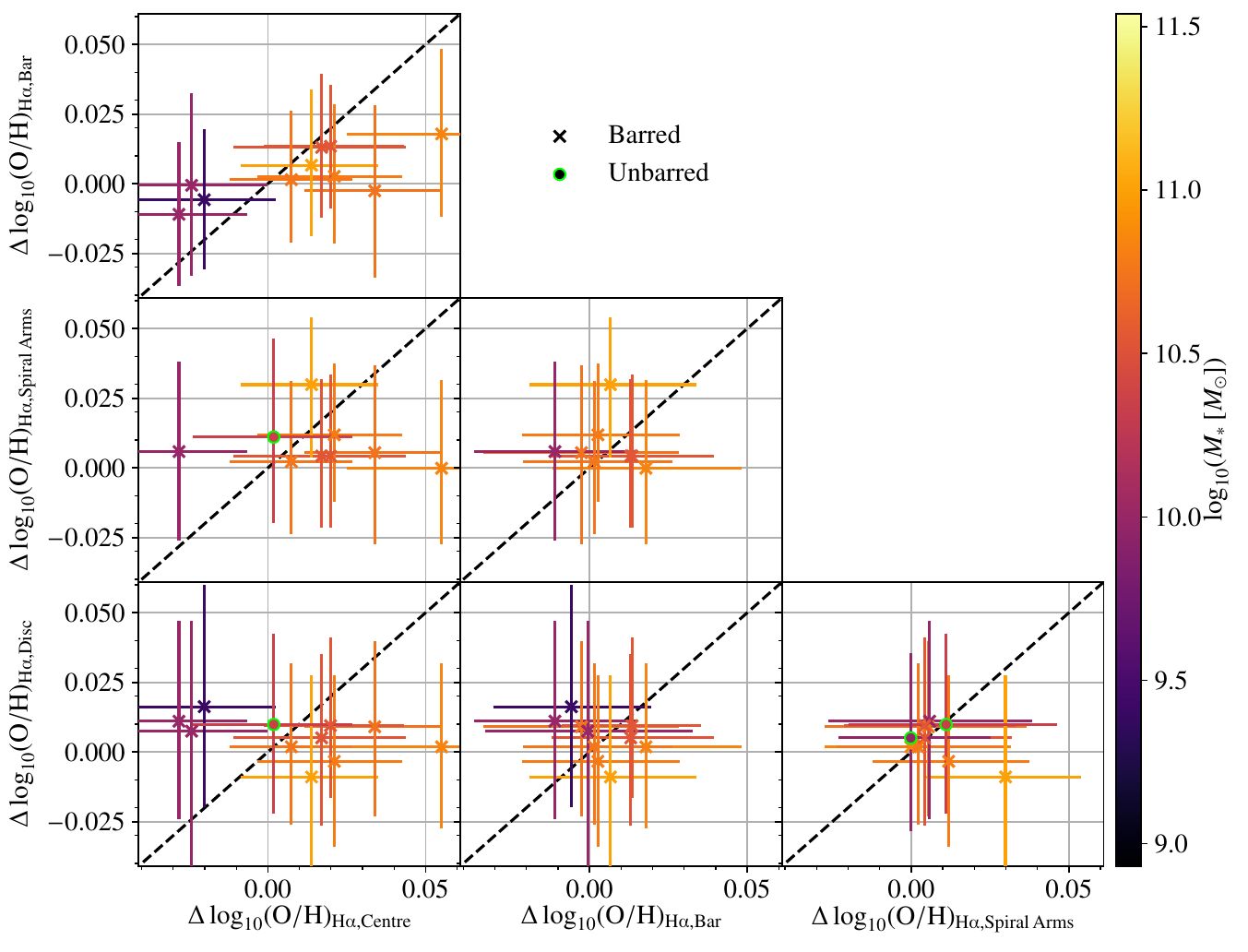}
\caption{$\Delta\oh_{\ha}$ for various environments for the \ngal\ galaxies with significant higher-order variations. The error bars represent the 16th and 84th percentiles of the KDE distribution, and the point marks the median. In many cases, the spread in values in galaxy centres is very narrow, and so the error bars are comparable to the size of the point. In each case, the point is coloured by the stellar mass of the galaxy, and we indicated barred galaxies with crosses, and unbarred with green-outlined circles. The ${1\!:\!1}$ line is shown as a dashed-black line.} 
\label{fig:abundance_offsets}
\end{figure*}

With an interpolated metallicity map, we can now study how the metallicity varies in different environments of the galaxy. Abundance variations have been predicted from numerical galaxy simulations \citep[e.g.][]{2013DiMatteo, 2016Grand}, but observational works show conflicting results. Some studies have reported no significant variation between the spiral arm and inter-arm regions \citep[e.g.][]{1996MartinBelley,2002CedresCepa, 2016Kreckel}, whilst others have found evidence that metallicities are enhanced in spiral arms with respect to the inter-arm region \citep[e.g.][]{2018Sakhibov,2020SanchezMenguiano}. These variations may be spatially localised, perhaps only occurring in one of the spiral arms \citep{2019Kreckel}, or could vary radially \citep[and peaking at co-rotation;][]{2019Spitoni}. With these maps, we can now address if (and where) metallicities are enhanced with a large, homogeneous sample of high-resolution metallicity maps. 

To do this, we take the environmental masks from \cite{2021Querejeta}. To avoid being too granular in defining environments, we use the `simple' masks as defined in that work, which divide galaxies up into `centres', `bars', `spiral arms', and `discs' (which includes the outer disc, the entire disc of galaxies without strong spiral arms, as well as inter-arm and inter-bar regions). We reproject these masks onto our radially-subtracted metallicity maps, and calculate a Kernel Density Estimate (KDE) of the radially subtracted metallicities ($\Delta\oh$; i.e.\ the top right panel of Fig.~\ref{fig:gpr_fit}) for each pixel grouped by its environment. Because some environments may have more sparse metallicity measurements (see e.g. along the bar in NGC~3627 in Fig. \ref{fig:gpr_fit}), we also perturb the values by their associated errors before calculating the KDE. The error reported by the GPR represents the sparsity of measurements along with the uncertainty of each given measurement, and so this perturbation will account for environments with fewer metallicity measurements (e.g. along bars). For the KDE, we use a \cite{Silverman} bandwidth (see also this reference for an introduction to KDE). To avoid giving equal weight to more uncertain metallicity values in fainter regions, the KDE is calculated using weighting based on the $\ha$ flux, and refer to this distribution as $\Delta\oh_{\ha}$. Because the bar may also promote efficient mixing and reduces local abundance variations \citep{2013DiMatteo}, we split the centres of galaxies up further into a barred and unbarred sample. For our \ngal\ galaxies, this is shown in Fig.~\ref{fig:abundance_offsets_kde}. The distributions for all environments look reasonably Gaussian and (except the centres of barred galaxies, which tend to be slightly higher) peak around a $\Delta\oh_{\ha}$ of $0.01$~dex, indicating that, when averaged across the entire sample, there are no clear overall enhancements within each environment (i.e. that the amplitude of the fitted GPR averages to zero over these larger scales, and not that the residual metallicity distribution is necessarily flat within each environment).

However, by averaging across the whole sample, we may be washing out variations between galaxies. We therefore repeat this exercise for each galaxy individually, calculating the KDE distribution of the radially subtracted metallicities for each environment, and weighting each measurement by the $\ha$ intensity. We use the median of these distributions (i.e.\ the median of each environmental distribution in Fig.~\ref{fig:abundance_offsets_kde} for each galaxy) to indicate the average abundance offset in that particular environment, and the 16th and 84th percentiles to show the spread of values. We show this for each galaxy in Fig.~\ref{fig:abundance_offsets}. In this formalism, points that lie above the ${1\!:\!1}$ line have enhanced abundances in the $y$-axis environment with respect to the $x$-axis environment (and vice versa). We see that generally the $\Delta\oh_{\ha}$ values are consistent with~0, showing little indication of abundance variations within galaxies. In particular, the spiral arm and disc $\Delta\oh_{\ha}$ values are very close to zero (apart from NGC~1365 towards the bottom right of this subplot), contrary to the results of \cite{2018Sakhibov} and \cite{2020SanchezMenguiano}, using similar definitions for the spiral arms and disc. Note that although we include the outer disc, the FOV of the MUSE observations typically does not extend far into this regime. The only environment where we see clear abundance variations are in the centres of galaxies, which typically show enhanced metallicity with respect to the rest of the galactic environments, even after subtraction of the bulk radial metallicity gradient. This effect becomes more pronounced with stellar mass (the correlation between $\Delta\oh_{\ha, \rm Centre}$ with $\log_{10} (M_\ast$) is $\tau=0.42\pm0.29$), with more massive galaxies having a more metal-enriched centre with respect to the rest of the galaxy than expected from their overall (linear) radial metallicity gradient. The fact that we see this is not surprising, as the metallicity gradient typically appears to be steeper towards the centres of galaxies \citep[e.g.][]{2014Sanchez, 2016SanchezMenguiano}, and can also be seen in the lower than expected CO-to-H$_2$ conversion factor seen in the centres of galaxies \citep{2013Sandstrom}. 

\section{What drives variation in the ISM mixing scale?}\label{sec:global_correlations}

\begin{figure*}
\includegraphics[width=1.8\columnwidth]{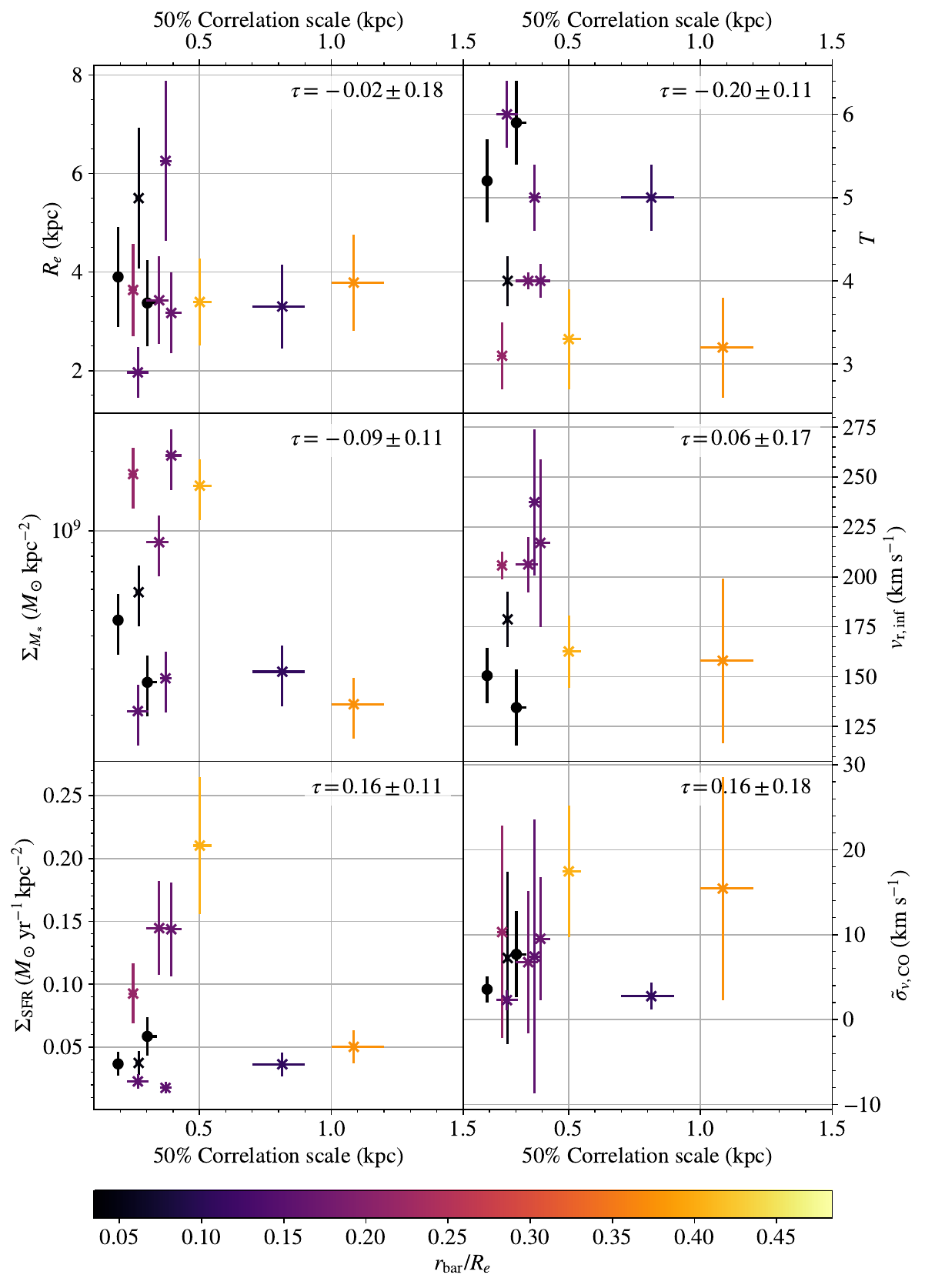}
\caption{From top left to bottom right, the dependence on the 50~per~cent two-point correlation scale with $R_{e}$, total stellar surface density (stellar mass normalised by $\pi R_{e}^2$), total star formation rate surface density, Hubble type~$T$, asymptotic rotation velocity, and CO~intensity-weighted CO~velocity dispersion. In each case, Kendall's $\tau$ correlation coefficient is shown in the top right. Barred galaxies are indicated as crosses, and coloured according to their (normalised) bar length. Unbarred galaxies are black circles.} 
\label{fig:scale_length_global_quantities}
\end{figure*}

We may expect the global properties of the galaxy to encode some information about their mixing scale. For instance, more massive or earlier type galaxies may have more homogeneous metal distributions, and so may display a larger scale length \citep{2021Li}, perhaps because the majority of enrichment happened much earlier in the lifecycle of the galaxy. Turbulence in the ISM can be injected via cosmological gas accretion \citep{2010KlessenHennebelle} or increased star formation \citep{2018Krumholz}, and this increased turbulence may increase metal mixing. We show the relationship between the two-point correlation scale and measures of the evolutionary state, star formation activity, and gas turbulence in Fig.~\ref{fig:scale_length_global_quantities}. As in Sect.~\ref{sec:two_point_corr}, we exclude the anomalously high value for NGC~1365. We normalise mass quantities by $\pi R_{e}^2$ (the effective radius that contains half the stellar mass of the galaxy), to give a disc-averaged surface density. 

We see a number of correlations with these quantities.
Firstly, galaxies with a higher Hubble type~$T$ typically have a lower 50~per~cent correlation scale. This indicates that earlier-type galaxies have a more homogeneous metal distribution. However, this is not echoed in the galaxy size, stellar mass surface density or total mass \citep[from the asymptotic rotation velocity calculated by][]{2020Lang}. In \cite{2021Li}, trends were found with the galaxy size, and we do not find this here (either with $R_e$, or $r_{25}$ as used in their work). However, we do find the same trends with the stellar mass surface density as in \cite{2021Li}. We also find a trend with the star formation rate surface density, unlike this earlier work. These results seem to indicate the correlation scales seen are driven by a combination of factors -- first, the stage of a galaxy in its lifecycle, from the morphological type, and secondly its current star formation activity.

We also study how the correlation scale depends on the gas velocity dispersion. The stochastically forced diffusion model of \cite{2018KrumholzTing} suggest that the correlation scale should strongly correlate with the gas velocity dispersion. We therefore compare our correlation scale to the CO velocity dispersions derived from the PHANGS--ALMA data \citep[][we use the strict moment~0 map for intensity, and moment~2 map for velocity dispersion]{2021Leroy}. We calculate a CO~intensity-weighted CO~velocity dispersion measured at ${\sim}100$~pc resolution across the whole CO~map as an average measure of the cold gas velocity dispersion, and the 16th and 84th percentiles as a measure of the spread. Given the spectral resolution of the MUSE instrument, in many cases the $\ha$ velocity dispersion is barely resolved and so we do not include it here, but with the higher spectral resolution of the ${\sim}2.5~\mathrm{km\,s^{-1}}$ ALMA data we expect this to be less of an issue. This is shown in the bottom right panel of Fig.~\ref{fig:scale_length_global_quantities}. We see no clear trend with gas velocity dispersion, which was also seen in \cite{2021Li} for a larger sample of galaxies. We also note that repeating this exercise using the coarser velocity resolution MUSE data gives identical results, and replacing 50~per~cent correlation scales with $\sigma_l$ show the same trends. These results are thus in contradiction with the \cite{2018KrumholzTing} model. However, as we are averaging over the entire galaxy here, we may be missing some local effects. Indeed, \cite{2020Kreckel} found that in $2$~kpc annular rings, the two-point correlation scales are strongly correlated with the gas velocity dispersion within that ring. Our analysis may therefore be washing out these local effects. Our results suggest the single largest global drivers of mixing are the morphological type~$T$ of the galaxy, and star formation activity, with mass, size, and gas turbulence playing little role in predicting the global metal mixing efficiency.

\section{Conclusions}\label{sec:conclusions}

In this work, we have mapped the two-dimensional variations of metals across the discs of 19 nearby galaxies. We have done this by calculating a `per-pixel' metallicity at a common worst resolution of $120$~pc, subtracting the dominant linear radial metallicity gradient, and then performing an interpolation using Gaussian Process Regression. The key parameter that we extract from this fitting is $\sigma_l$, a characteristic kernel length scale that is indicative of the distance over which neighbouring measurements are highly correlated. We have performed a McJack procedure to correct these kernel length scales for incomplete coverage, and establish whether they are statistically significant. We find that, in our sample, \ngal\ of our 19 galaxies show significant higher-order variations. Those that do not, typically have many fewer metallicity measurements, and so this may be limited to the lack of \hii\ regions in the galaxy, and not reflect a real lack of higher-order variations in the galaxy. Our measured kernel length scales range from $1.5$ to $30$~kpc, with a median value of $15$~kpc. We compared our kernel length scales to the 50~per~cent correlation scale from a two-point correlation function, and find them to be related, although the kernel length scales measured in this work are typically around an order of magnitude larger. The two-point correlation functions measured from our maps are similar to those measured for \hii\ region catalogues, indicating the GPR fitting is sensitive to the same small-scale features as the two-point correlation function.

With these \ngal\ galaxies, we have investigated how the residual (i.e. radially-subtracted) metal enrichment varies with galactic environment (e.g.\ spiral arms, bars). We have divided each galaxy up into centres, bars, spiral arms, and discs using the environmental masks from \cite{2021Querejeta}. We see no clear signs of enrichment in any particular environment (e.g. spiral arms, disc) when taking all galaxies as a whole, but see that centres are typically enriched (up to $0.05$~dex higher, dependent on the total stellar mass of the galaxy) with respect to the linear radial metallicity gradient. We find no evidence that spiral arms are enriched compared to the disc, unlike recent work from \cite{2020SanchezMenguiano}. Abundance variations within a particular environment that we fit with the GPR typically average out when considering the entire environment. However, azimuthal variations \citep{2019Kreckel}, or fluctuations that vary radially \citep{2019Spitoni} would be missed by our analysis.

We have also looked at how the 50~per~cent correlation scale from the two-point correlation function varies with different global galaxy parameters. Higher star formation rate surface density and lower Hubble type~$T$ have larger correlation scales. This may indicate that galaxies that are more evolved and with higher levels of star formation activity have mixed (or are mixing) their metals more efficiently. Unlike predictions from the model of \cite{2018KrumholzTing}, we find no significant correlation between the scale length and global gas velocity dispersion, in agreement with the findings of \cite{2021Li}.

Whilst generally a second-order effect compared to the dominant radial metallicity gradient in galaxies, higher-order variations appear to be ubiquitous and non-negligible (with variations of up to $0.05$~dex from the radial gradient, see Fig.~\ref{fig:abundance_offsets_kde}) in star-forming spiral galaxies. With advanced statistical techniques and high-quality data, we have demonstrated that it is possible to measure and map these variations in a statistically robust way. These kinds of models will be suitable for comparison to the outputs of simulations, where metallicities are known locally, and can be compared to the effects of large-scale dynamical processes, such as bar mixing \citep[e.g.][]{2016Grand}. Our interpolated metallicity maps also provide a minimally biased way to combine metallicities measurements from one observatory with observations at other wavelengths and resolutions, that probe different galactic properties, for resolved studies of, e.g., the dust-to-metals ratio \citep[e.g.][]{2019DeVis, 2021Chiang}, or studying the metallicity dependence on the CO-to-H$_2$ conversion factor, which is critical in obtaining reliable H$_2$ masses from CO data \citep[e.g.][]{2013Sandstrom}. In moving from one-dimensional, first-order radial gradients to two-dimensional, higher-order variations, this work provides a stepping stone towards more realistic models of metal variations within galaxies.

\section*{Acknowledgements}

This work has been carried out as part of the PHANGS collaboration. The authors would like to thank the anonymous reviewer, for comments and suggestions that have improved the quality of this paper.

This work has made use of {\tt AstroPy} \citep{2013Astropy,2018Astropy}, {\tt Matplotlib} \citep{2007Hunter}, {\tt NumPy} \citep{harris2020array}, {\tt SciPy} \citep{2020SciPy}, {\tt scikit-learn} \citep{scikit-learn}, and {\tt Seaborn} \citep{2017Waskom}.

This work is based on observations collected at the European Southern Observatory under ESO programmes 1100.B-0651, 095.C-0473, and 094.C-0623, 094.B-0321, 099.B0242, 0100.B-0116, 098.B-0551 and 097.B-0640. 

This paper makes use of the following ALMA data, which have been processed as part of the PHANGS--ALMA \mbox{CO($J=2{-}1$)}\ survey: \\

\noindent ADS/JAO.ALMA\#2012.1.00650.S, \linebreak % (N628/M74)
ADS/JAO.ALMA\#2013.1.00803.S, \linebreak % (N5128/CenA)
ADS/JAO.ALMA\#2013.1.01161.S, \linebreak % (N1365 + N5236/M83)
ADS/JAO.ALMA\#2015.1.00121.S, \linebreak % (N5236/M83)
ADS/JAO.ALMA\#2015.1.00782.S, \linebreak % (N1313 + N7793)
ADS/JAO.ALMA\#2015.1.00925.S, \linebreak % (pilot low mass)
ADS/JAO.ALMA\#2015.1.00956.S, \linebreak % (pilot high mass)
ADS/JAO.ALMA\#2016.1.00386.S, \linebreak % (N5236/M83)
ADS/JAO.ALMA\#2017.1.00392.S, \linebreak % (low mass follow-up)
ADS/JAO.ALMA\#2017.1.00766.S, \linebreak % (early-type)
ADS/JAO.ALMA\#2017.1.00886.L, \linebreak % (large program)
ADS/JAO.ALMA\#2018.1.01321.S, \linebreak % (N253, N300, Circinus)
ADS/JAO.ALMA\#2018.1.01651.S, \linebreak % (main sample follow-up)
ADS/JAO.ALMA\#2018.A.00062.S, \linebreak % (ACA-only nearby)
ADS/JAO.ALMA\#2019.1.01235.S, \linebreak % (local sample follow up)
ADS/JAO.ALMA\#2019.2.00129.S, \linebreak % (N1068)

ALMA is a partnership of ESO (representing its member states), NSF (USA), and NINS (Japan), together with NRC (Canada), NSC and ASIAA (Taiwan), and KASI (Republic of Korea), in cooperation with the Republic of Chile. The Joint ALMA Observatory is operated by ESO, AUI/NRAO, and NAOJ. The National Radio Astronomy Observatory is a facility of the National Science Foundation operated under cooperative agreement by Associated Universities, Inc.

TGW, FS, H-AP, ES acknowledge funding from the European Research Council (ERC) under the European Union’s Horizon 2020 research and innovation programme (grant agreement No. 694343). KK gratefully acknowledges funding from the German Research Foundation (DFG) in the form of an Emmy Noether Research Group (grant number KR4598/2-1, PI Kreckel). FB acknowledges funding from the European Research Council (ERC) under the European Union’s Horizon 2020 research and innovation programme (grant agreement No.726384/Empire). MB acknowledges FONDECYT regular grant 1170618. MC and JMDK gratefully acknowledge funding from the Deutsche Forschungsgemeinschaft (DFG, German Research Foundation) through an Emmy Noether Research Group (grant number KR4801/1-1), as well as from the European Research Council (ERC) under the European Union's Horizon 2020 research and innovation programme via the ERC Starting Grant MUSTANG (grant agreement number 714907). EC acknowledges support from ANID project Basal AFB-170002. The work of AKL was partially supported by the National Science Foundation (NSF) under Grants No.1615105, and 1653300. ER acknowledges the support of the Natural Sciences and Engineering Research Council of Canada (NSERC), funding reference number RGPIN-2017-03987. SCOG, EJW and RSK acknowledge support from the DFG via the collaborative research center (SFB 881, Project-ID 138713538) “The Milky Way System” (subprojects A1, B1, B2, B8 and P2). SCOG and RSK additionally acknowledge support from the Heidelberg Cluster of Excellence ``STRUCTURES'' in the framework of Germany’s Excellence Strategy (grant EXC-2181/1, Project-ID 390900948) and from the ERC via the ERC Synergy Grant ``ECOGAL'' (grant 855130). MQ acknowledges support from the research project PID2019-106027GA-C44 from the Spanish Ministerio de Ciencia e Innovaci{\'o}n. KS acknowledges funding support from National Science Foundation Award No. 1816462. EWK acknowledges support from the Smithsonian Institution as a Submillimeter Array (SMA) Fellow.

%%%%%%%%%%%%%%%%%%%%%%%%%%%%%%%%%%%%%%%%%%%%%%%%%%
\section*{Data Availability}

The MUSE data used in this paper are presented in \cite{2021Emsellem}. The ALMA data are presented in \cite{2021Leroy}, and available online at \url{https://www.canfar.net/storage/list/phangs/RELEASES/PHANGS-ALMA/}. The code used in this paper, along with metallicity maps (and associated error maps) are available at \url{https://github.com/thomaswilliamsastro/metallicity_gpr}.

% The inclusion of a Data Availability Statement is a requirement for articles published in MNRAS. Data Availability Statements provide a standardised format for readers to understand the availability of data underlying the research results described in the article. The statement may refer to original data generated in the course of the study or to third-party data analysed in the article. The statement should describe and provide means of access, where possible, by linking to the data or providing the required accession numbers for the relevant databases or DOIs.



\bibliographystyle{mnras}
\bibliography{bibliography}

%\begin{thebibliography}{99}
%\end{thebibliography}

%%%%%%%%%%%%%%%%%%%%%%%%%%%%%%%%%%%%%%%%%%%%%%%%%%

%%%%%%%%%%%%%%%%% APPENDICES %%%%%%%%%%%%%%%%%%%%%

\appendix

\section{GPR fits for all PHANGS-MUSE galaxies}\label{app:all_fit_overviews}

Here, we show the equivalent of Fig. \ref{fig:gpr_fit} for all 19 galaxies in our sample.

\input{fit_overview}

\section{\texorpdfstring{\hii}{HII}\ region two-point correlation function}\label{app:two_point_hii}

We revisit the two-point correlation function calculation from \cite{2020Kreckel}, based on the \hii\ region positions and metallicities,  for the full sample of 19 galaxies.  This analysis makes use of the \hii\ region catalogue presented in \cite{2021Santoro} and metallicity measurements presented in Groves et al. (in prep.), and is thus an update on the values previously published for eight galaxies in \cite{2020Kreckel} that were based on the \hii\ region catalogue in \cite{2019Kreckel}. In addition to more than doubling the galaxy sample, the updated \hii\ region catalogue is based on a new data reduction, which includes a more accurate sky subtraction among other improvements \citep{2021Emsellem}. The \hii\ region identification algorithm has also been refined. As a result of these improvements, the newer catalogues contain 20-30\% more \hii\ region detections per galaxy. 

We select as \hii\ regions all nebulae that meet the following criteria:
\begin{itemize}
    \item S/N~$>$~5 in all strong lines ($\hb$, $\oiii$, $\nii$, $\ha$, $\sii$) that are used as diagnostics for the ionization source or the metallicity calculation;
    \item line ratios consistent with photoionzation in the BPT diagrams using the \cite{2003Kauffmann} diagnostic in the $\nii$ diagram the \cite{2001Kewley} diagnostic in the $\sii$ diagram;
    \item spatial separation by more than the PSF from any bright foreground stars or the field edges;
    \item $\ha$ velocity dispersion $<$ $60~{\rm km\,s^{-1}}$, to remove supernova remnant contaminants and spurious fits.
\end{itemize}
The resulting catalogue consists of a total of 23,436 regions, and identifies between 476 and 2355 \hii\ regions per galaxy. 

Metallicities are computed by adopting the \cite{2016PilyuginGrebel} S\nobreakdash-\hspace{0pt}calibration, and removing a linear radial gradient as fit in \cite{2021Santoro}.  In computing the \hii\ region separations, distances are deprojected within each galaxy and computed assuming the position angle and inclination, as listed in Table \ref{tab:galaxy_params}. 

\begin{figure*}
    \centering
    \includegraphics[width=\textwidth]{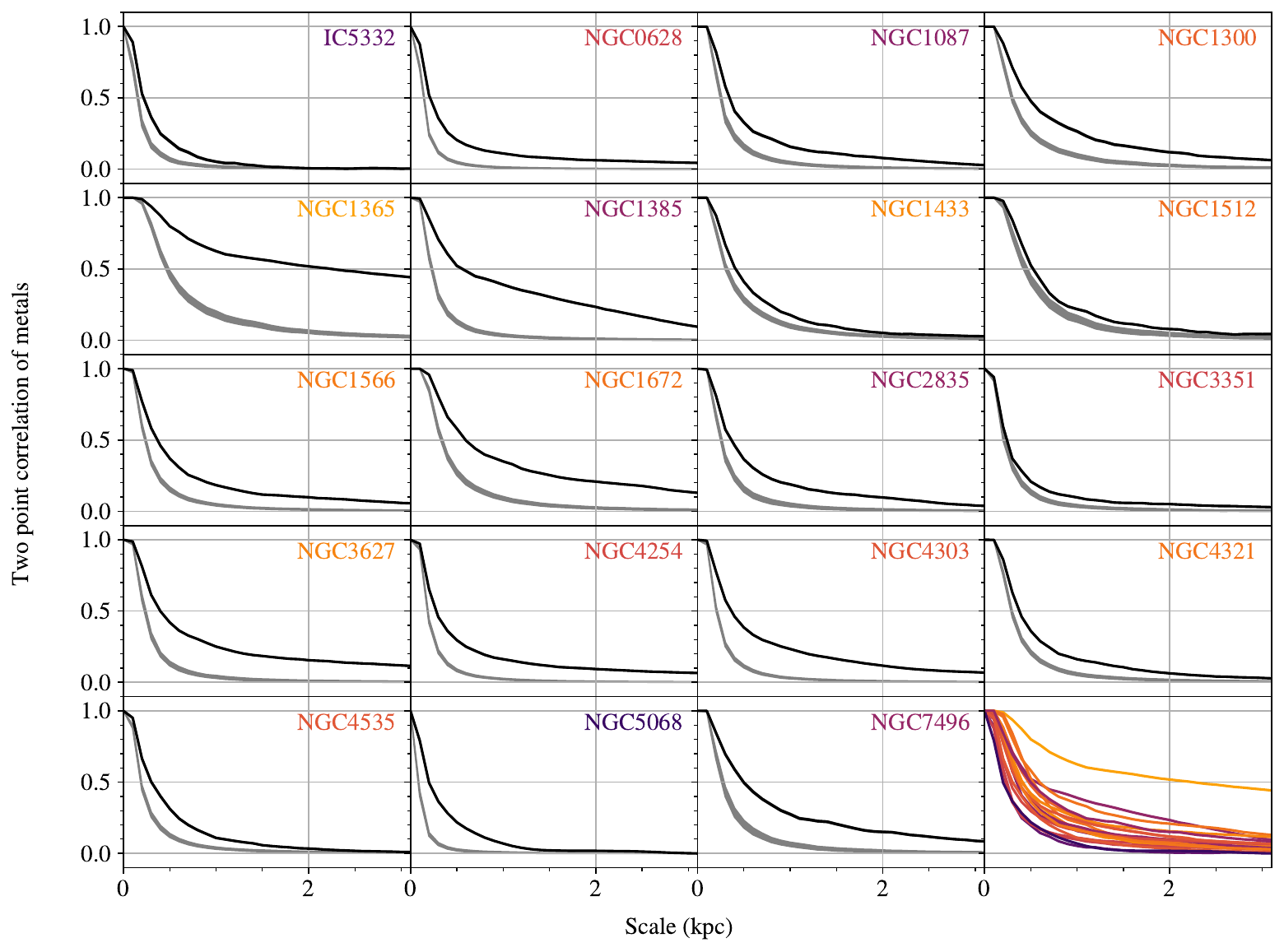}
    \caption{Two point correlation of metals for each of the 19 PHANGS--MUSE galaxies. Galaxies colours are ordered by their total stellar mass, from low (purple) to high (orange). We estimate the uncertainty in our measured function by performing 100 random samples of our uncertainties in $\met$, and repeating our analysis.  A band is plotted covering the $1\sigma$ distribution at each spatial scale, though it is generally not visible as it is thinner than the width of the line ($<$10pc). To determine the significance of our two-point correlation functions, we assume the null hypothesis (that all \hii\ regions are perfectly uncorrelated) by randomly shuffling (for 100 actualizations) all measurements of $\met$ across each galaxy and repeating our analysis.  This is shown by the grey bands.}
    \label{fig:2pt}
\end{figure*}

Within each galaxy, we calculate the two point correlation of metals ($\xi$) as a function of spatial scale ($r$) using Eq. \ref{eq:two_point_corr}. The resulting two-point correlation functions are shown in Figure \ref{fig:2pt}.  We estimate the uncertainty in our measured function by performing 100 random samples of our uncertainties in $\met$, and repeating our analysis.  The $1\sigma$ distribution is determined at each spatial scale, though it is generally thinner than the line drawn ($<$10pc). To determine the significance of our two-point correlation functions, we assume the null hypothesis (that all \hii\ regions are perfectly uncorrelated) by randomly shuffling (for 100 actualizations) all measurements of $\met$ across each galaxy and repeating our analysis.  This is shown by the grey bands, and in all galaxies the two-point correlation functions are measured at $>$2 $\sigma$ out to kpc scales. 
%A 100\% correlation is expected at scales of r=0, as each \hii\ region correlates perfectly with itself. The shape of the correlation function traces the scale over which the metal field retains some level of homogeneity. In the following analysis, we parameterize these curves by the scale at which the metallicities remain correlated at the 30\% or 50\% level.  

\begin{figure*}
    \centering
    \includegraphics[width=\textwidth]{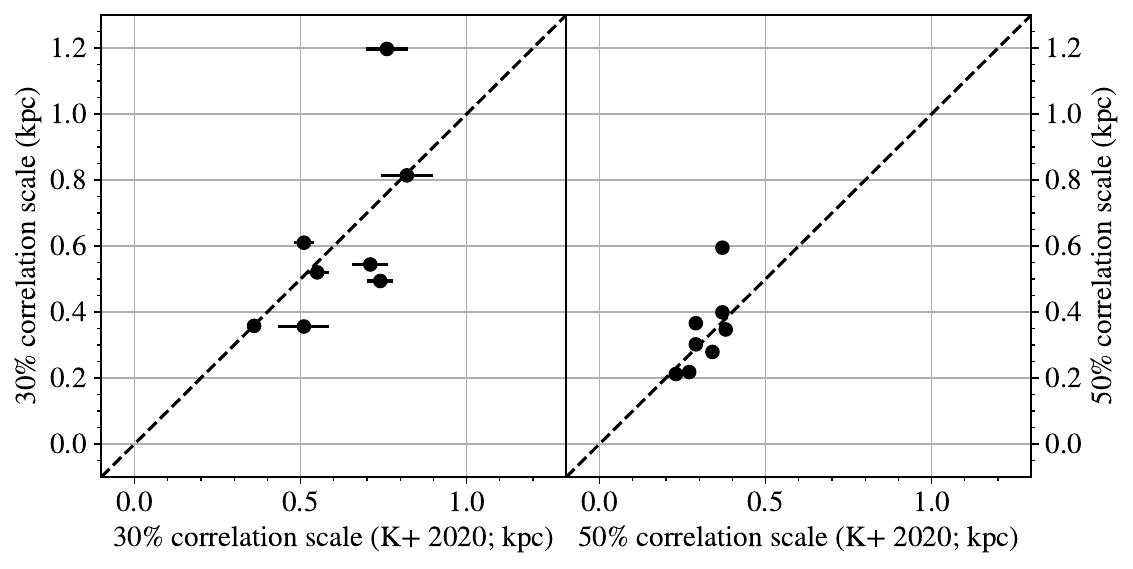}
    \caption{Comparison of the 30~per~cent and 50~per~cent correlation scales for the eight galaxies in \citet{2020Kreckel} with the measurements provided in this work (y-axis). Error bars reflect the $1\sigma$ uncertainties, which in most cases are smaller than the points. NGC~1672 shows the most discrepant values between the two \hii\ region catalogues, with the remaining galaxies showing good agreement.}
    \label{fig:KK20_compare}
\end{figure*}

We quantify the 30~per~cent and 50~per~cent correlation scales for each galaxy, along with the $1\sigma$ uncertainties, with all values listed in Table \ref{tab:2pt}. A direct comparison with the values reported in \cite{2020Kreckel} are shown in Figure \ref{fig:KK20_compare}, and reveal very good systematic agreement for the eight overlapping galaxies.

NGC~1365 shows remarkably high correlations out to the largest scale we test (5kpc). NGC~3351 and NGC~1512 show the least significant difference from the null hypothesis, with only $\sim$2$\sigma$ difference at the 30~per~cent and 50~per~cent correlation scales.  Measurements for all other galaxies are detected at $>$6$\sigma$ levels. As shown in the bottom right corner of Figure \ref{fig:2pt}, only a weak trend with galaxy stellar mass (indicated by the line colours) is apparent. 

\begin{table*}
\caption{Correlation scales measured for the 19 galaxies. }
\label{tab:2pt}
\begin{tabular}{lcc}
\hline\hline
Galaxy & 50~per~cent correlation & 30~per~cent correlation \\
 & scale (pc) & scale (pc) \\
\hline
IC5332 &  218$\pm$  4 &  356$\pm$  5 \\
NGC0628 &  212$\pm$  1 &  358$\pm$  2 \\
NGC1087 &  347$\pm$  3 &  544$\pm$  8 \\
NGC1300 &  476$\pm$  5 &  870$\pm$ 16 \\
NGC1365 & 2247$\pm$ 29 & $>$5000 \\
NGC1385 &  561$\pm$ 10 & 1529$\pm$ 17 \\
NGC1433 &  412$\pm$  3 &  666$\pm$  7 \\
NGC1512 &  525$\pm$  5 &  752$\pm$  7 \\
NGC1566 &  368$\pm$  3 &  615$\pm$  5 \\
NGC1672 &  595$\pm$  3 & 1197$\pm$  9 \\
NGC2835 &  366$\pm$  4 &  610$\pm$  8 \\
NGC3351 &  242$\pm$  2 &  380$\pm$  4 \\
NGC3627 &  399$\pm$  2 &  814$\pm$  6 \\
NGC4254 &  279$\pm$  1 &  494$\pm$  4 \\
NGC4303 &  367$\pm$  2 &  695$\pm$  5 \\
NGC4321 &  377$\pm$  1 &  585$\pm$  3 \\
NGC4535 &  302$\pm$  4 &  520$\pm$  5 \\
NGC5068 &  199$\pm$  1 &  379$\pm$  3 \\
NGC7496 &  497$\pm$  7 &  905$\pm$ 17 \\
\hline
\end{tabular}
\end{table*}

% \section{Scale length illustration}\label{app:scale_length_illustration}

% \begin{figure}
% \includegraphics[width=\columnwidth]{figs/scale_length_illustration.pdf}
% \caption{Sampled Gaussian process for a number of scale lengths (indicated in the top left of each subplot). With increasing scale length, the maps becoming increasingly smoother.} 
% \label{fig:scale_length_illustration}
% \end{figure}

% To illustrate the effects of increasing $\sigma_l$ on maps, we generate a number of 2D fields using different values for $\sigma_l$ (not that here $\nu=1.5$, as used in our fitting). We use a grid of $100\times100$ points, and sample values from a Gaussian process. This is shown in Fig.~\ref{fig:scale_length_illustration}. As $\sigma_l$ increases, the maps become increasingly smoother and more featureless as points are correlated over larger distances. However, $\sigma_l$ is not trivially a ``distance between peak/trough''.

% If you want to present additional material which would interrupt the flow of the main paper,
% it can be placed in an Appendix which appears after the list of references.

%%%%%%%%%%%%%%%%%%%%%%%%%%%%%%%%%%%%%%%%%%%%%%%%%%

% Don't change these lines
\bsp	% typesetting comment
\label{lastpage}
\end{document}

%% file: galaxy_params.tex
\begin{table*}
\caption{Key parameters for the 19 galaxies. Galaxies with a bar are marked with an asterisk. Distances are taken from \citet{2021Anand}, PA and inclination from \citet{2020Lang}, and sizes, global stellar masses and global SFRs from \citet{2021Leroy}. The uncertainty in stellar mass is dominated by calibration uncertainty, so is always 0.11~dex for these galaxies. We also include the number of pixels in our convolved and regridded maps, and the number of pixels that have metallicity measurements after various cuts (see Sect. \ref{sec:metallicity_calculation}).}
\label{tab:galaxy_params}
\begin{tabular}{lccccccccr}
\hline\hline
Galaxy & Dist (Mpc) & PA (deg) & $i$ (deg) & $r_{25}$ (arcmin) & $R_e$ (arcmin) &  $\log_{10}(M_\ast [M_\odot])$ & SFR ($M_\odot\,{\rm yr^{-1}}$) & $N_{\rm pix}$ & $N_Z$ \\
\hline
IC5332 & 9.01 & 74.4 & 26.90 & 3.03 & 1.38 & $9.67\pm0.11$ & $0.41\pm0.11$ & 8239 & 362\\
NGC0628 & 9.84 & 20.7 & 8.90 & 4.94 & 1.36 & $10.34\pm0.11$ & $1.75\pm0.45$ & 23028 & 2193\\
NGC1087$^\ast$ & 15.85 & 359.1 & 42.90 & 1.49 & 0.70 & $9.93\pm0.11$ & $1.31\pm0.34$ & 32592 & 3133\\
NGC1300$^\ast$ & 18.99 & 278.0 & 31.80 & 2.97 & 1.18 & $10.62\pm0.11$ & $1.17\pm0.30$ & 94322 & 1587\\
NGC1365$^\ast$ & 19.57 & 201.1 & 55.40 & 6.01 & 0.49 & $10.99\pm0.11$ & $16.90\pm4.38$ & 108960 & 4541\\
NGC1385 & 17.22 & 181.3 & 44.00 & 1.70 & 0.67 & $9.98\pm0.11$ & $2.09\pm0.54$ & 32169 & 3634\\
NGC1433$^\ast$ & 18.63 & 199.7 & 28.60 & 3.10 & 0.79 & $10.87\pm0.11$ & $1.13\pm0.29$ & 113824 & 1434\\
NGC1512$^\ast$ & 18.83 & 261.9 & 42.50 & 4.22 & 0.87 & $10.71\pm0.11$ & $1.28\pm0.33$ & 70615 & 1457\\
NGC1566$^\ast$ & 17.69 & 214.7 & 29.50 & 3.61 & 0.62 & $10.78\pm0.11$ & $4.54\pm1.17$ & 54142 & 5044\\
NGC1672$^\ast$ & 19.4 & 134.3 & 42.60 & 3.08 & 0.60 & $10.73\pm0.11$ & $7.60\pm1.97$ & 66011 & 4806\\
NGC2835$^\ast$ & 12.22 & 1.0 & 41.30 & 3.21 & 0.93 & $10.00\pm0.11$ & $1.24\pm0.32$ & 22212 & 2234\\
NGC3351$^\ast$ & 9.96 & 193.2 & 45.10 & 3.61 & 1.05 & $10.36\pm0.11$ & $1.32\pm0.34$ & 18746 & 364\\
NGC3627$^\ast$ & 11.32 & 173.1 & 57.30 & 5.14 & 1.10 & $10.83\pm0.11$ & $3.84\pm1.00$ & 21696 & 3079\\
NGC4254 & 13.1 & 68.1 & 34.40 & 2.52 & 0.63 & $10.42\pm0.11$ & $3.07\pm0.79$ & 44203 & 7652\\
NGC4303$^\ast$ & 16.99 & 312.4 & 23.50 & 3.44 & 0.69 & $10.52\pm0.11$ & $5.33\pm1.38$ & 56444 & 8198\\
NGC4321$^\ast$ & 15.21 & 156.2 & 38.50 & 3.05 & 1.24 & $10.75\pm0.11$ & $3.56\pm0.92$ & 55510 & 4175\\
NGC4535$^\ast$ & 15.77 & 179.7 & 44.70 & 4.07 & 1.36 & $10.53\pm0.11$ & $2.16\pm0.56$ & 32153 & 1135\\
NGC5068$^\ast$ & 5.2 & 342.4 & 35.70 & 3.74 & 1.30 & $9.40\pm0.11$ & $0.28\pm0.07$ & 5224 & 1047\\
NGC7496$^\ast$ & 18.72 & 193.7 & 35.90 & 1.67 & 0.70 & $10.00\pm0.11$ & $2.26\pm0.59$ & 23140 & 1245\\
\hline
\end{tabular}
\end{table*}

%% file: scale_lengths_pg16_scal.tex
\begin{table}
\caption{Corrected scale lengths ($\sigma_l$), and 50~per~cent two-point correlation scale for the 12 galaxies with significant azimuthal variation.}
\label{tab:scale_lengths_corrected}
\begin{tabular}{lcr}
\hline\hline
Galaxy & $\sigma_l$ (kpc) & 50~per~cent Correlation Scale (kpc) \\
\hline
NGC0628 & $1.63^{+2.00}_{-1.19}$ & $0.19^{+0.00}_{-0.01}$\\
NGC1365 & $5.61^{+7.13}_{-4.11}$ & $4.12^{+0.08}_{-0.12}$\\
NGC1385 & $4.76^{+4.19}_{-3.28}$ & $0.30^{+0.04}_{-0.01}$\\
NGC1566 & $17.56^{+1.71}_{-2.47}$ & $0.39^{+0.04}_{-0.02}$\\
NGC1672 & $22.60^{+1.39}_{-2.26}$ & $0.50^{+0.05}_{-0.02}$\\
NGC2835 & $18.60^{+1.18}_{-2.36}$ & $0.81^{+0.09}_{-0.11}$\\
NGC3627 & $14.20^{+2.10}_{-3.49}$ & $0.25^{+0.02}_{-0.02}$\\
NGC4303 & $12.67^{+0.34}_{-0.41}$ & $0.35^{+0.04}_{-0.05}$\\
NGC4321 & $20.60^{+0.95}_{-2.36}$ & $0.27^{+0.01}_{-0.01}$\\
NGC4535 & $30.47^{+9.18}_{-11.34}$ & $0.37^{+0.02}_{-0.02}$\\
NGC5068 & $7.35^{+0.85}_{-1.32}$ & $0.27^{+0.04}_{-0.04}$\\
NGC7496 & $17.61^{+0.49}_{-0.84}$ & $1.08^{+0.12}_{-0.08}$\\
\hline
\end{tabular}
\end{table}

%% file: fit_overview.tex
\begin{figure*}
\includegraphics[width=2\columnwidth]{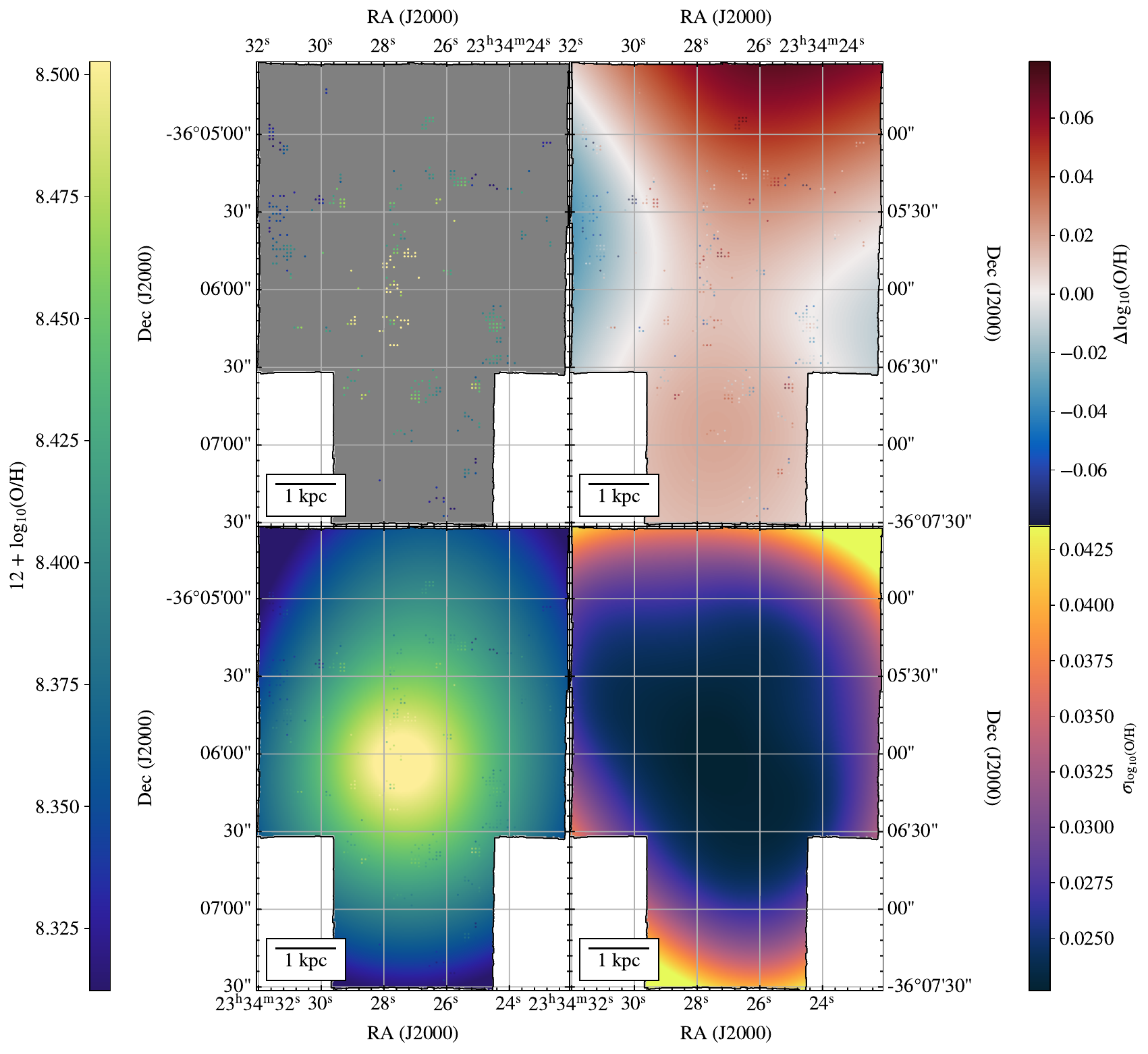}
\caption{As Fig. \ref{fig:gpr_fit}, but for IC5332}
\end{figure*}
\newpage
\begin{figure*}
\includegraphics[width=2\columnwidth]{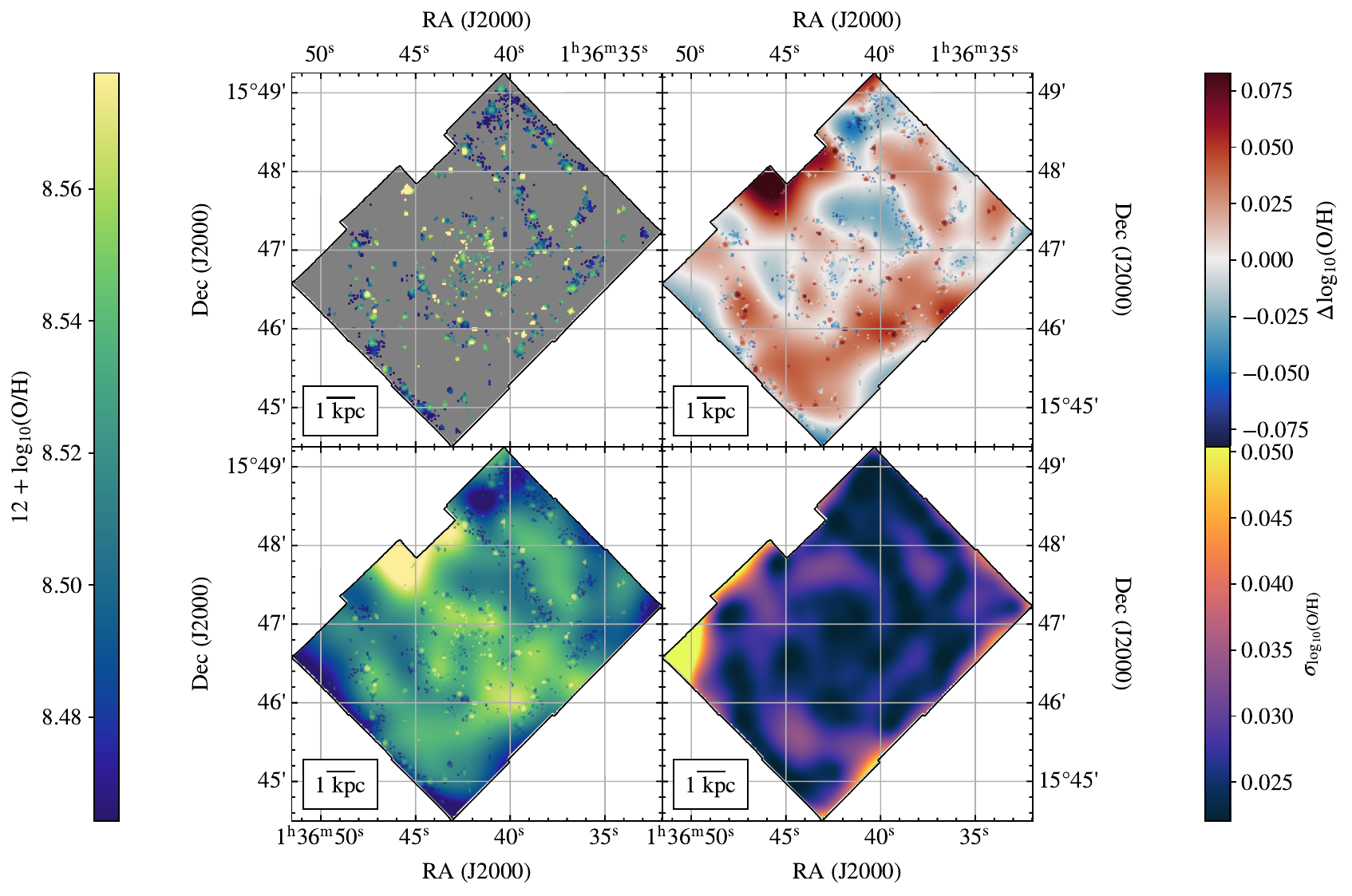}
\caption{As Fig. \ref{fig:gpr_fit}, but for NGC0628}
\end{figure*}
\newpage
\begin{figure*}
\includegraphics[width=2\columnwidth]{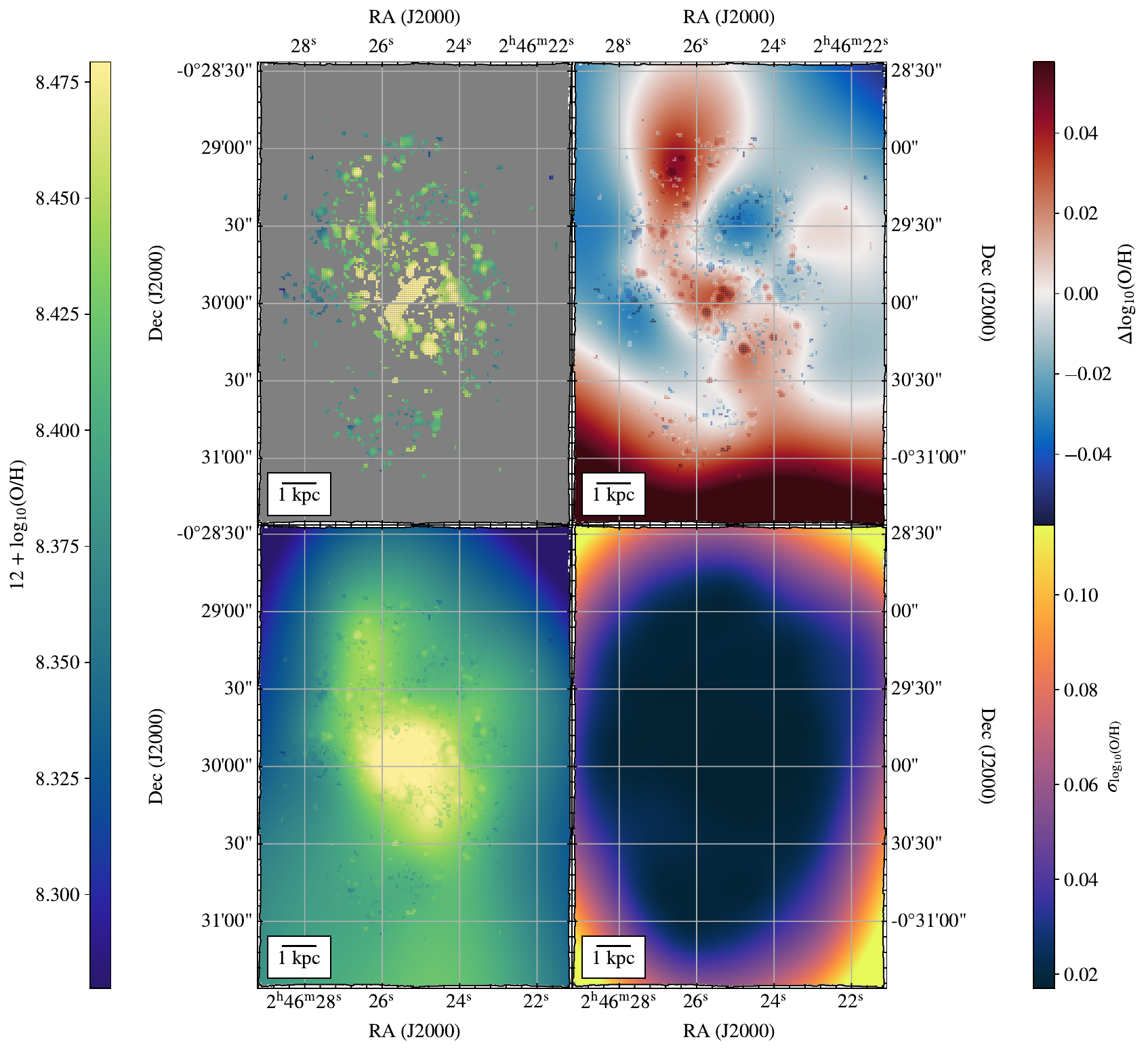}
\caption{As Fig. \ref{fig:gpr_fit}, but for NGC1087}
\end{figure*}
\newpage
\begin{figure*}
\includegraphics[width=2\columnwidth]{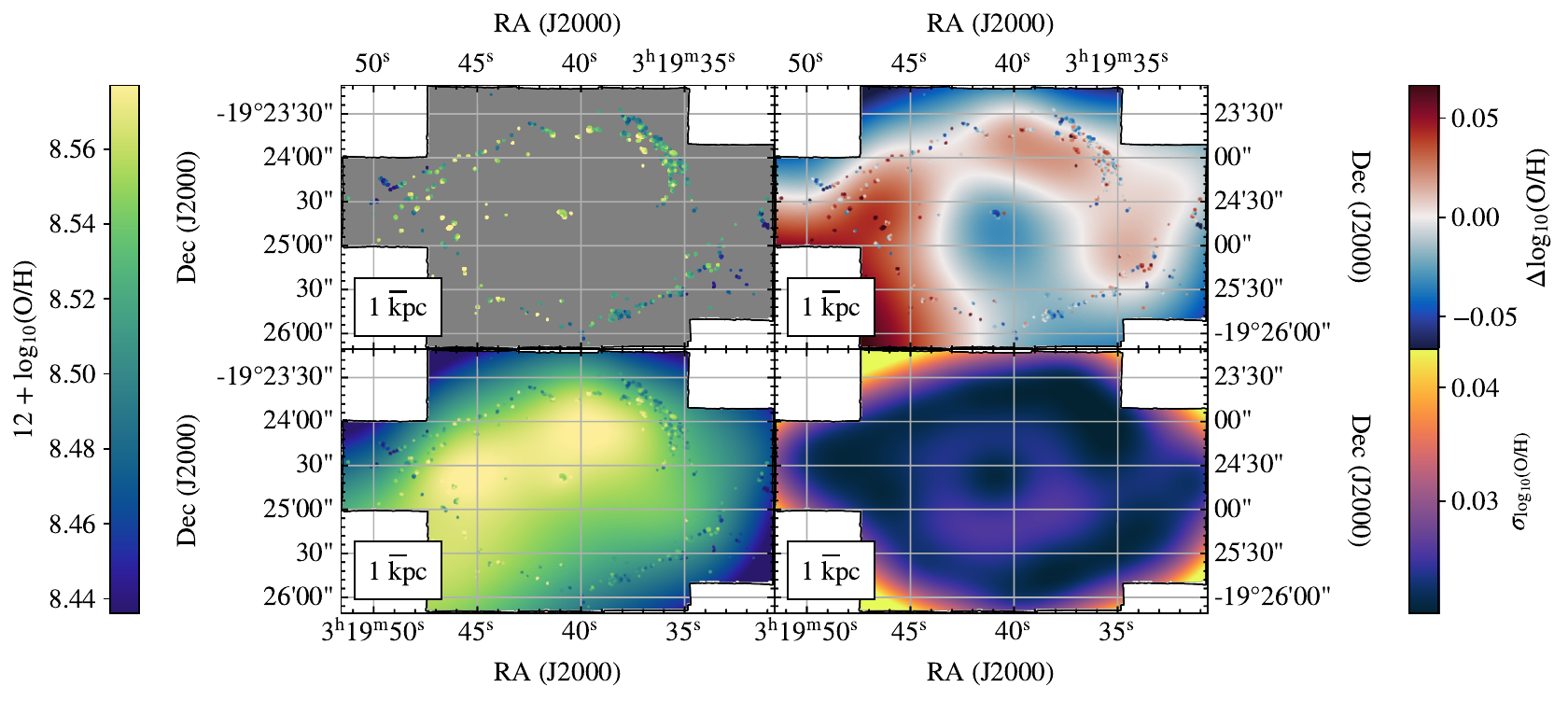}
\caption{As Fig. \ref{fig:gpr_fit}, but for NGC1300}
\end{figure*}
\newpage
\begin{figure*}
\includegraphics[width=2\columnwidth]{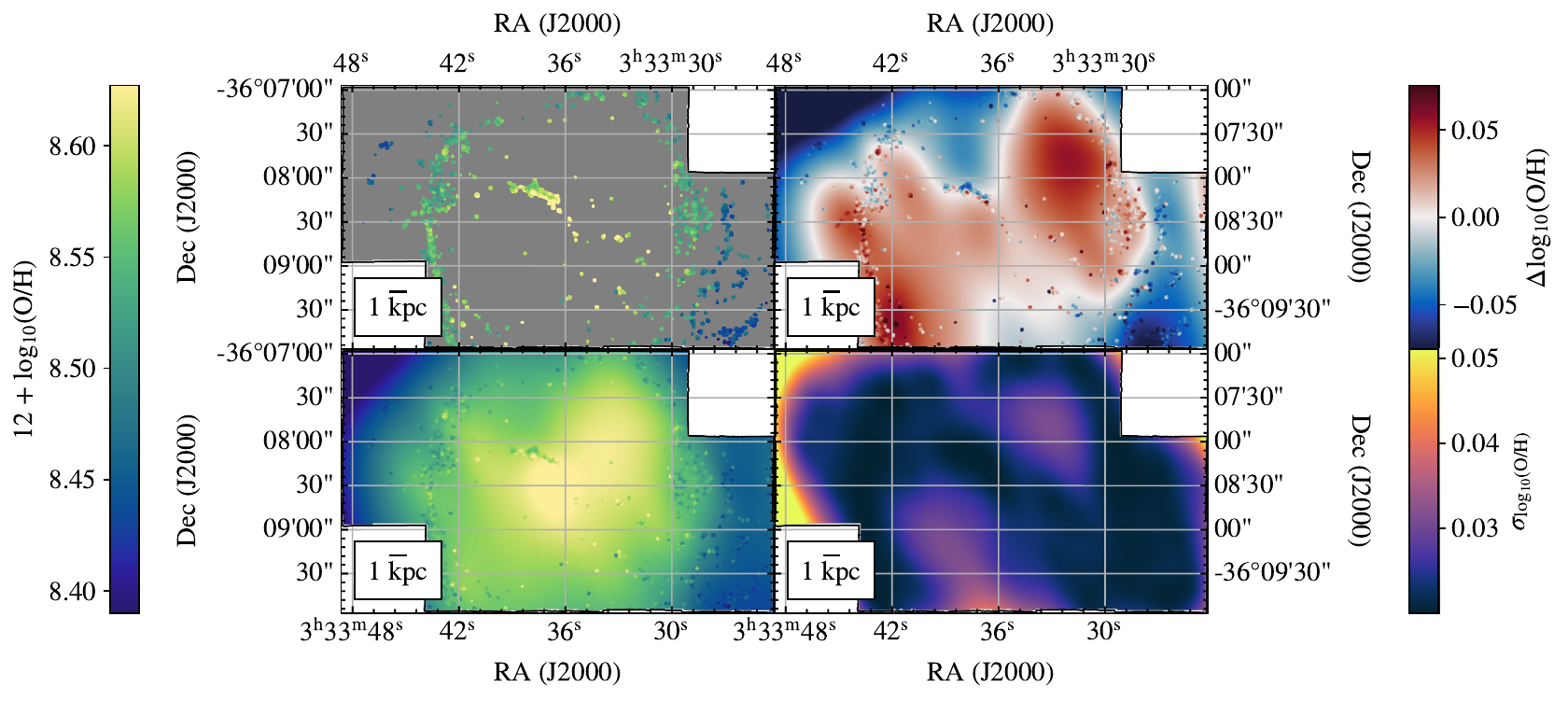}
\caption{As Fig. \ref{fig:gpr_fit}, but for NGC1365}
\end{figure*}
\newpage
\begin{figure*}
\includegraphics[width=2\columnwidth]{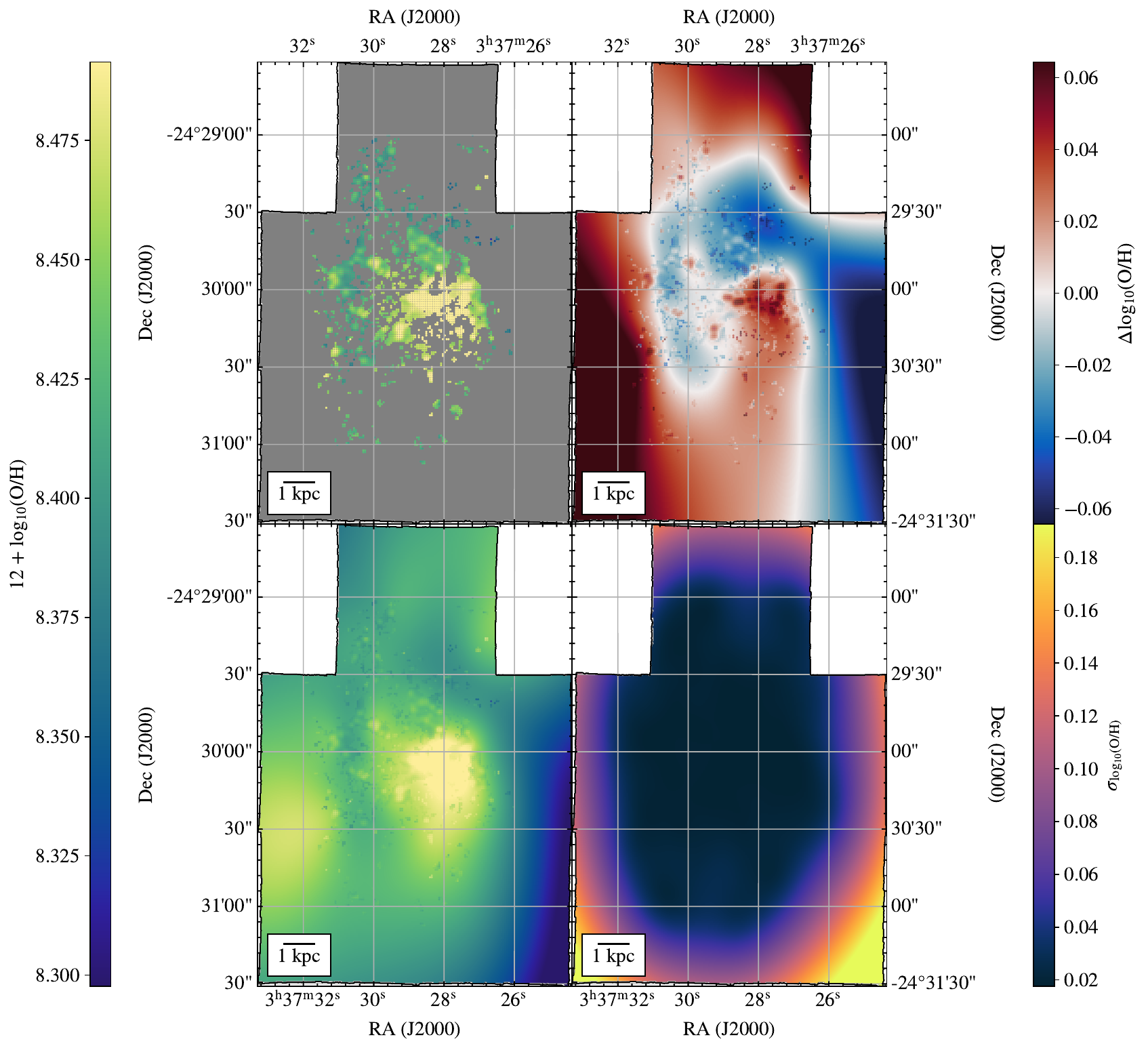}
\caption{As Fig. \ref{fig:gpr_fit}, but for NGC1385}
\end{figure*}
\newpage
\begin{figure*}
\includegraphics[width=2\columnwidth]{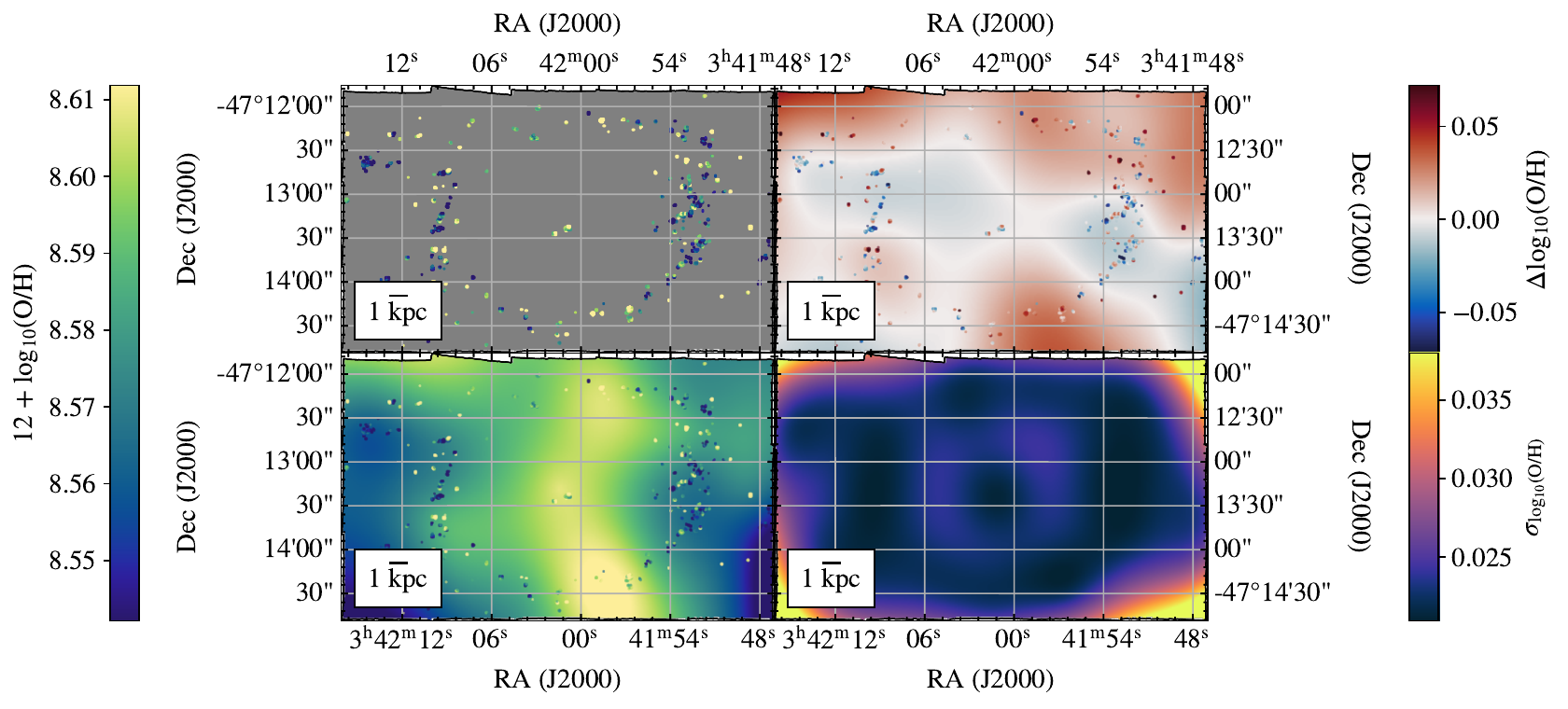}
\caption{As Fig. \ref{fig:gpr_fit}, but for NGC1433}
\end{figure*}
\newpage
\begin{figure*}
\includegraphics[width=2\columnwidth]{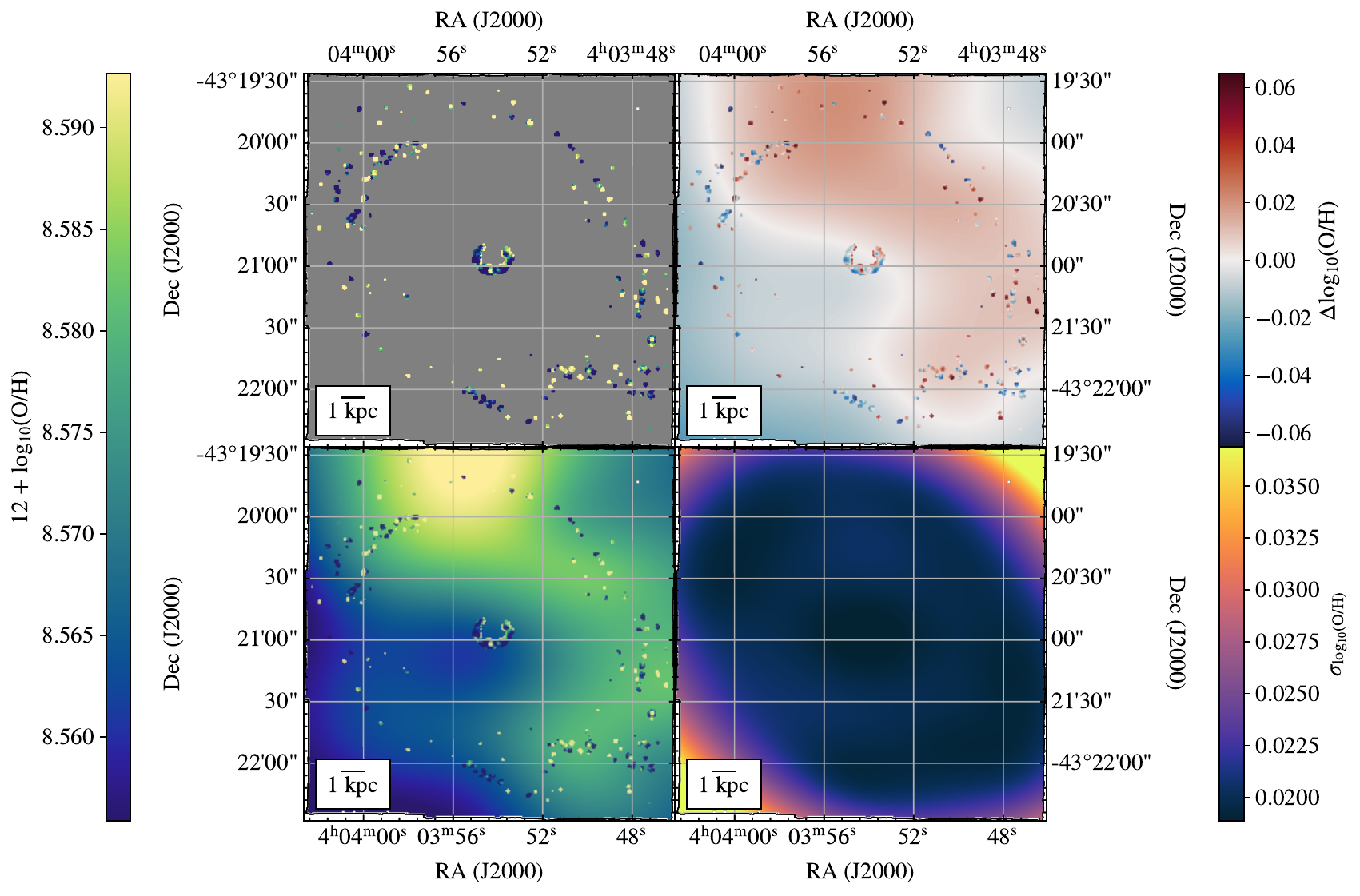}
\caption{As Fig. \ref{fig:gpr_fit}, but for NGC1512}
\end{figure*}
\newpage
\begin{figure*}
\includegraphics[width=2\columnwidth]{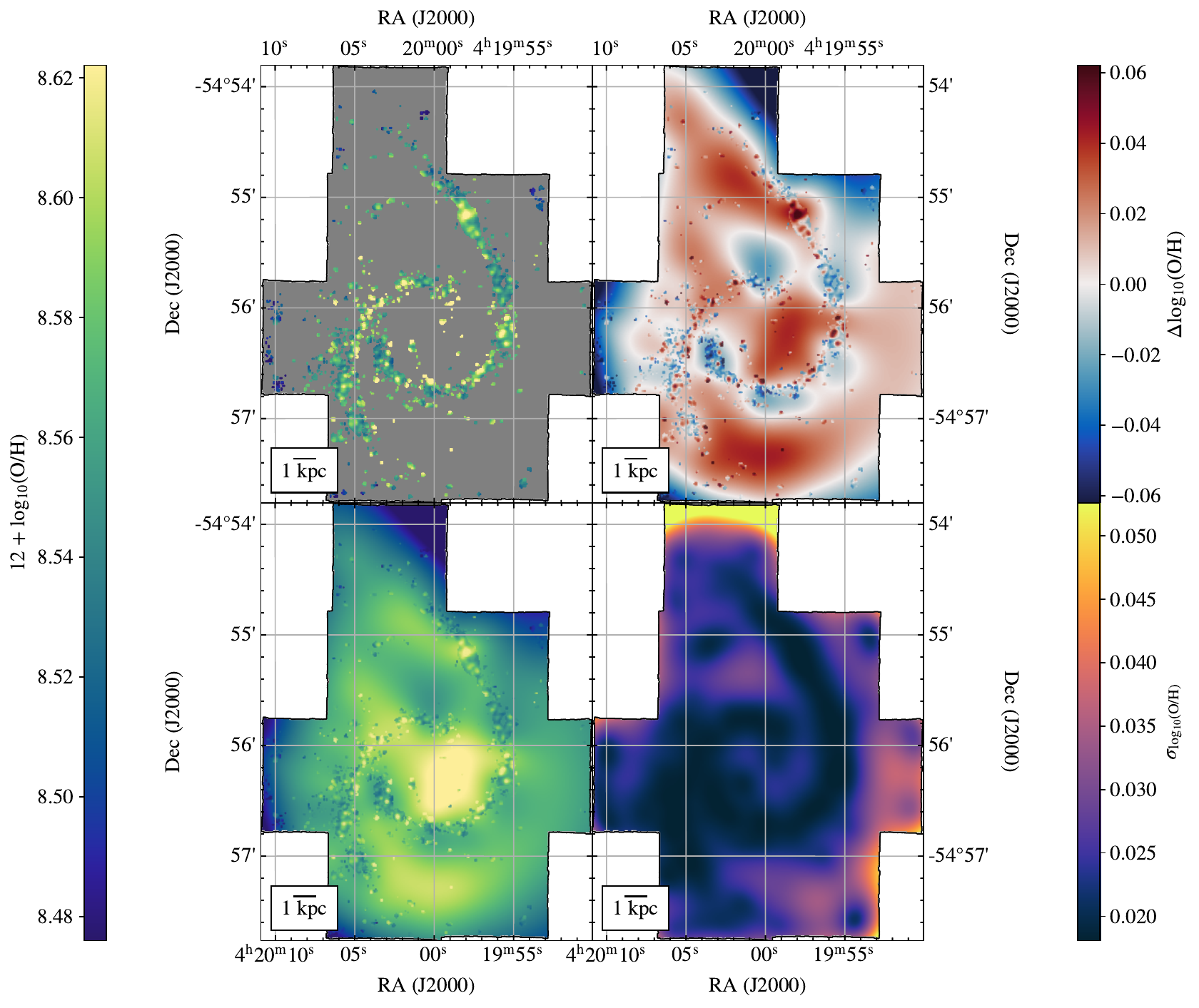}
\caption{As Fig. \ref{fig:gpr_fit}, but for NGC1566}
\end{figure*}
\newpage
\begin{figure*}
\includegraphics[width=2\columnwidth]{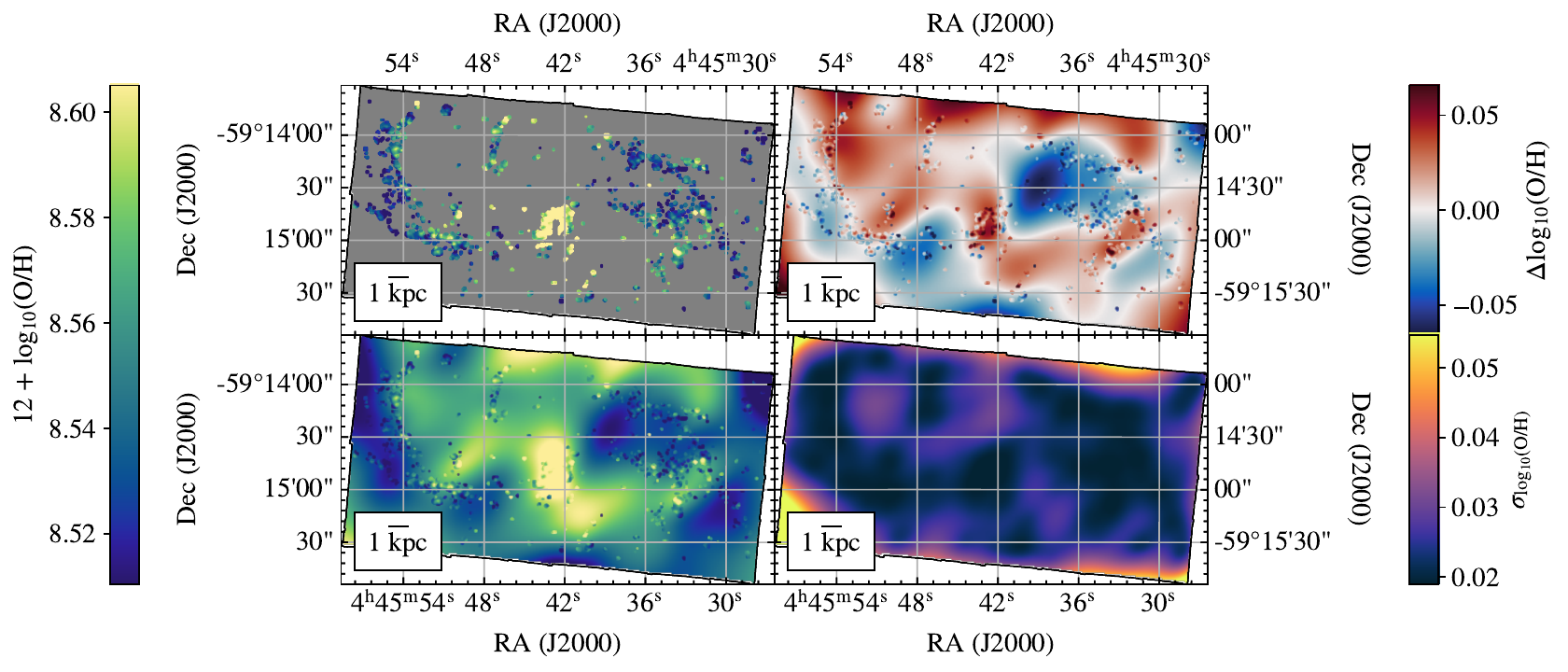}
\caption{As Fig. \ref{fig:gpr_fit}, but for NGC1672}
\end{figure*}
\newpage
\begin{figure*}
\includegraphics[width=2\columnwidth]{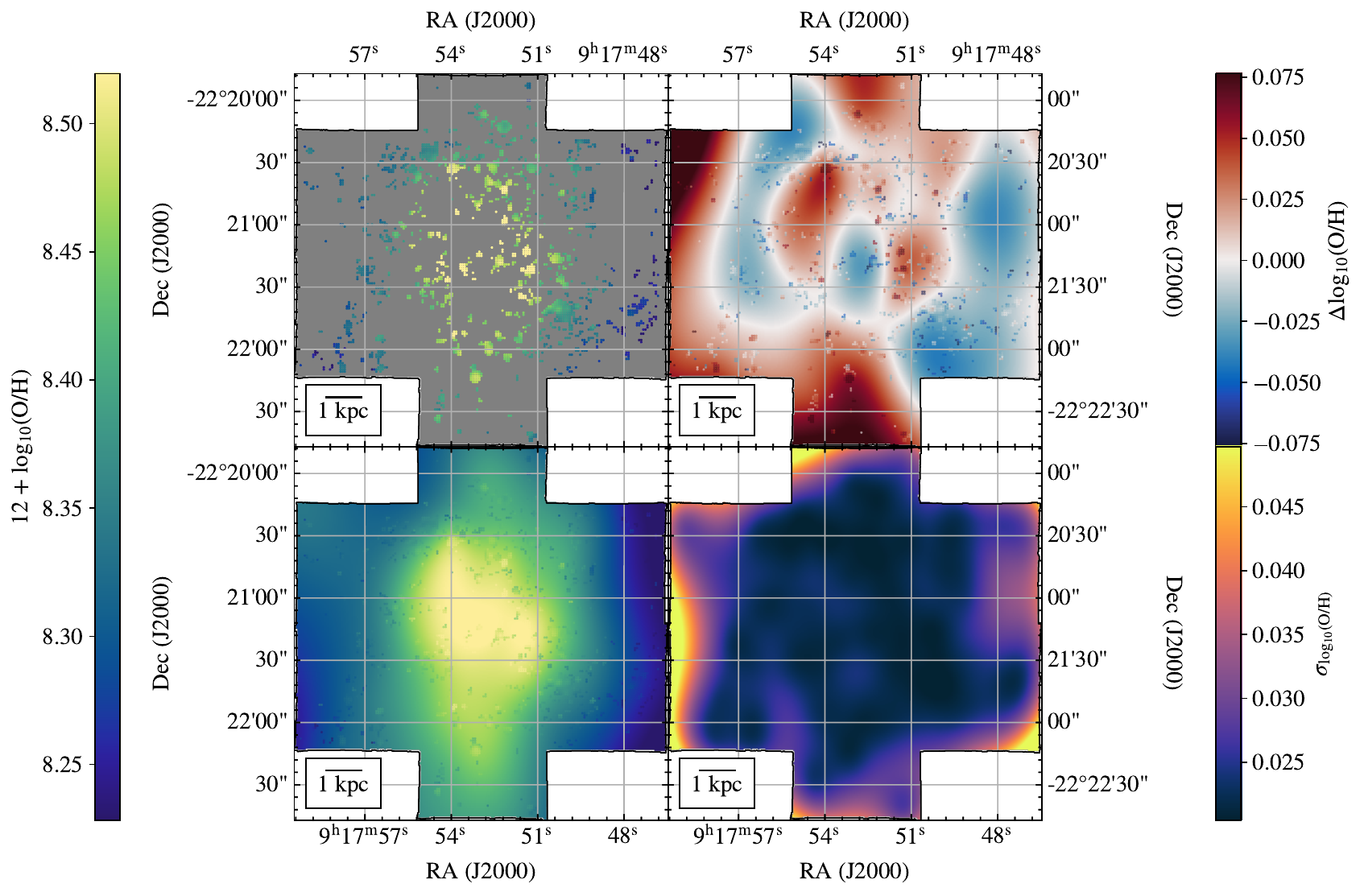}
\caption{As Fig. \ref{fig:gpr_fit}, but for NGC2835}
\end{figure*}
\newpage
\begin{figure*}
\includegraphics[width=2\columnwidth]{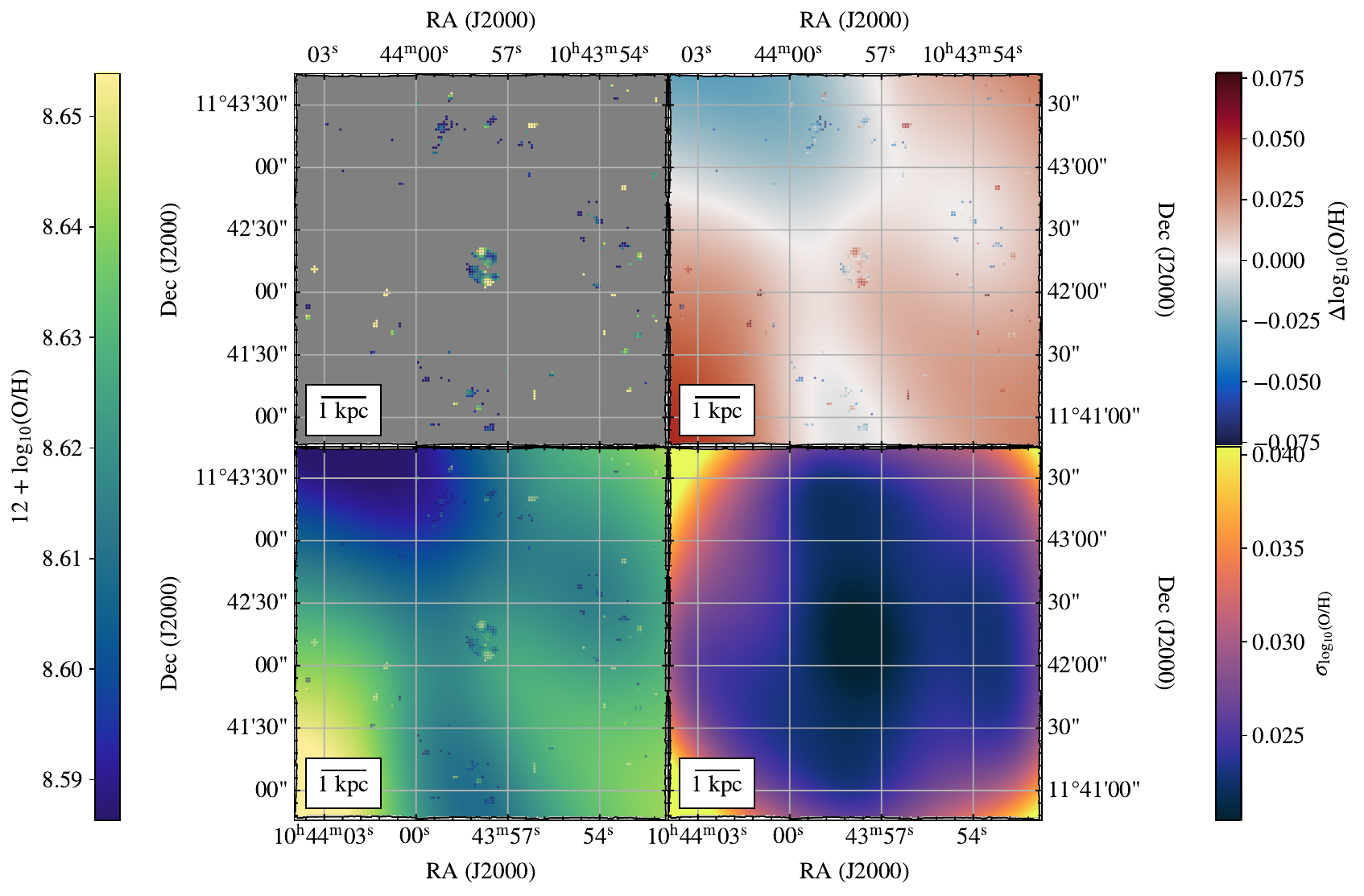}
\caption{As Fig. \ref{fig:gpr_fit}, but for NGC3351}
\end{figure*}
\newpage
\begin{figure*}
\includegraphics[width=2\columnwidth]{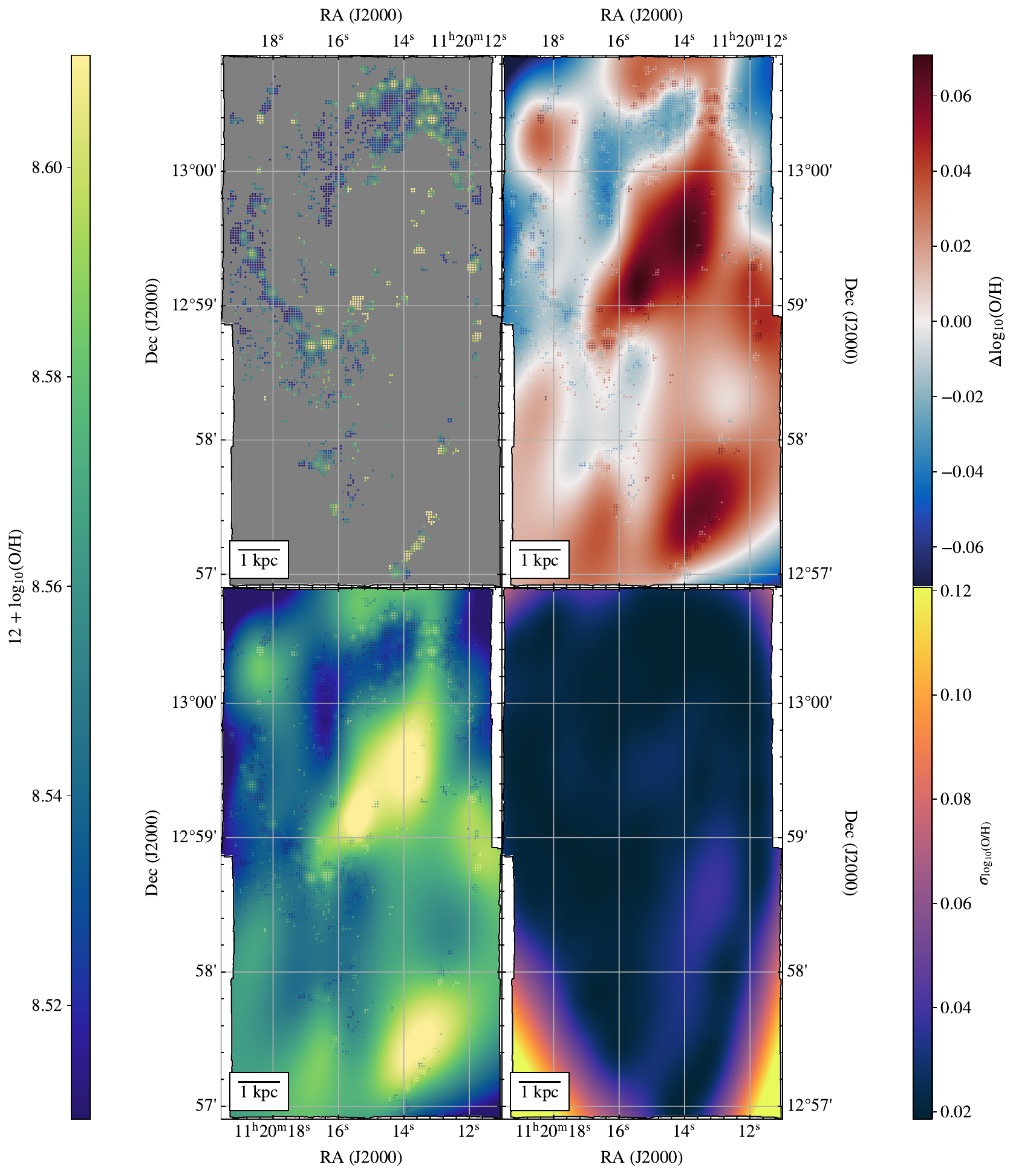}
\caption{As Fig. \ref{fig:gpr_fit}, but for NGC3627}
\end{figure*}
\newpage
\begin{figure*}
\includegraphics[width=2\columnwidth]{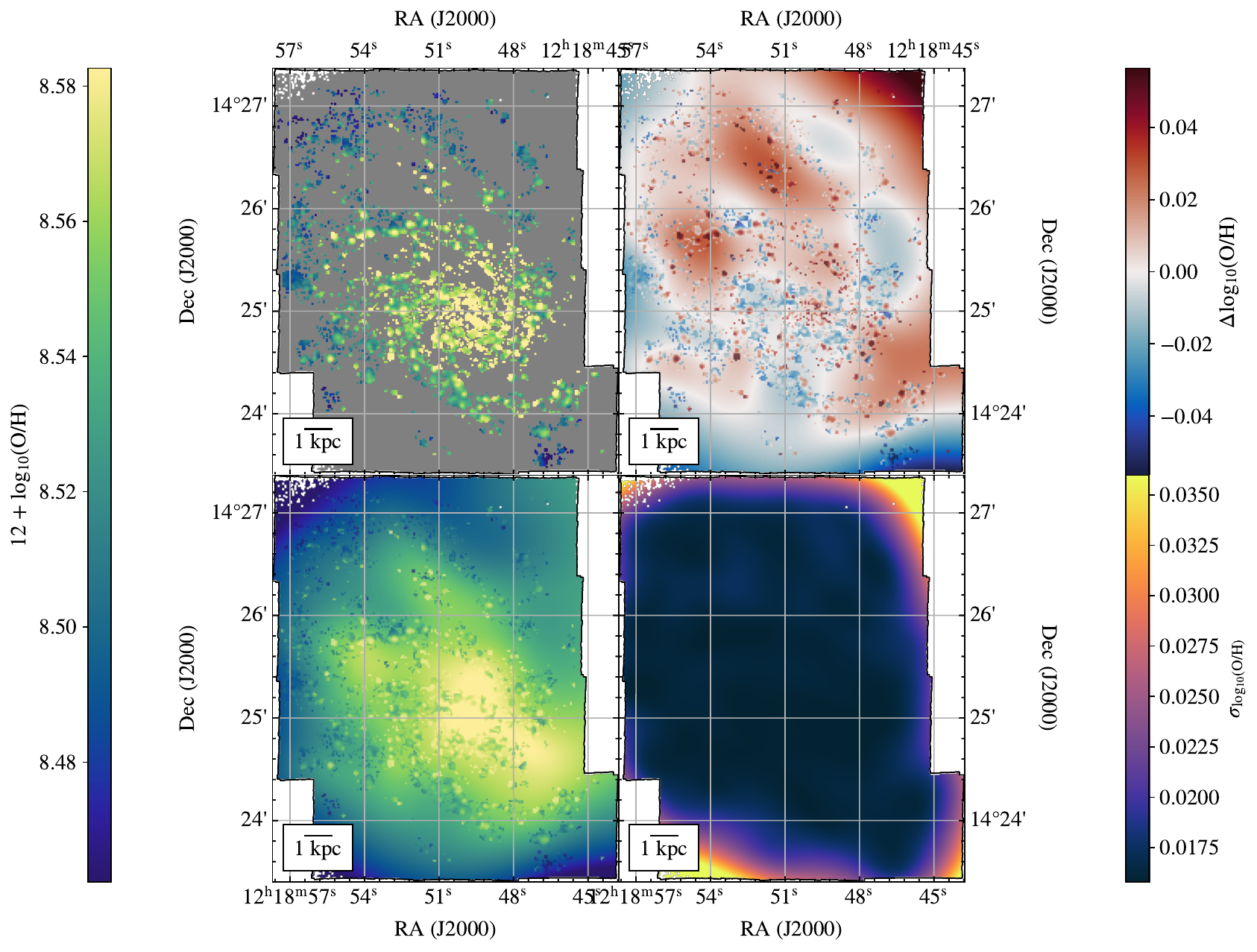}
\caption{As Fig. \ref{fig:gpr_fit}, but for NGC4254}
\end{figure*}
\newpage
\begin{figure*}
\includegraphics[width=2\columnwidth]{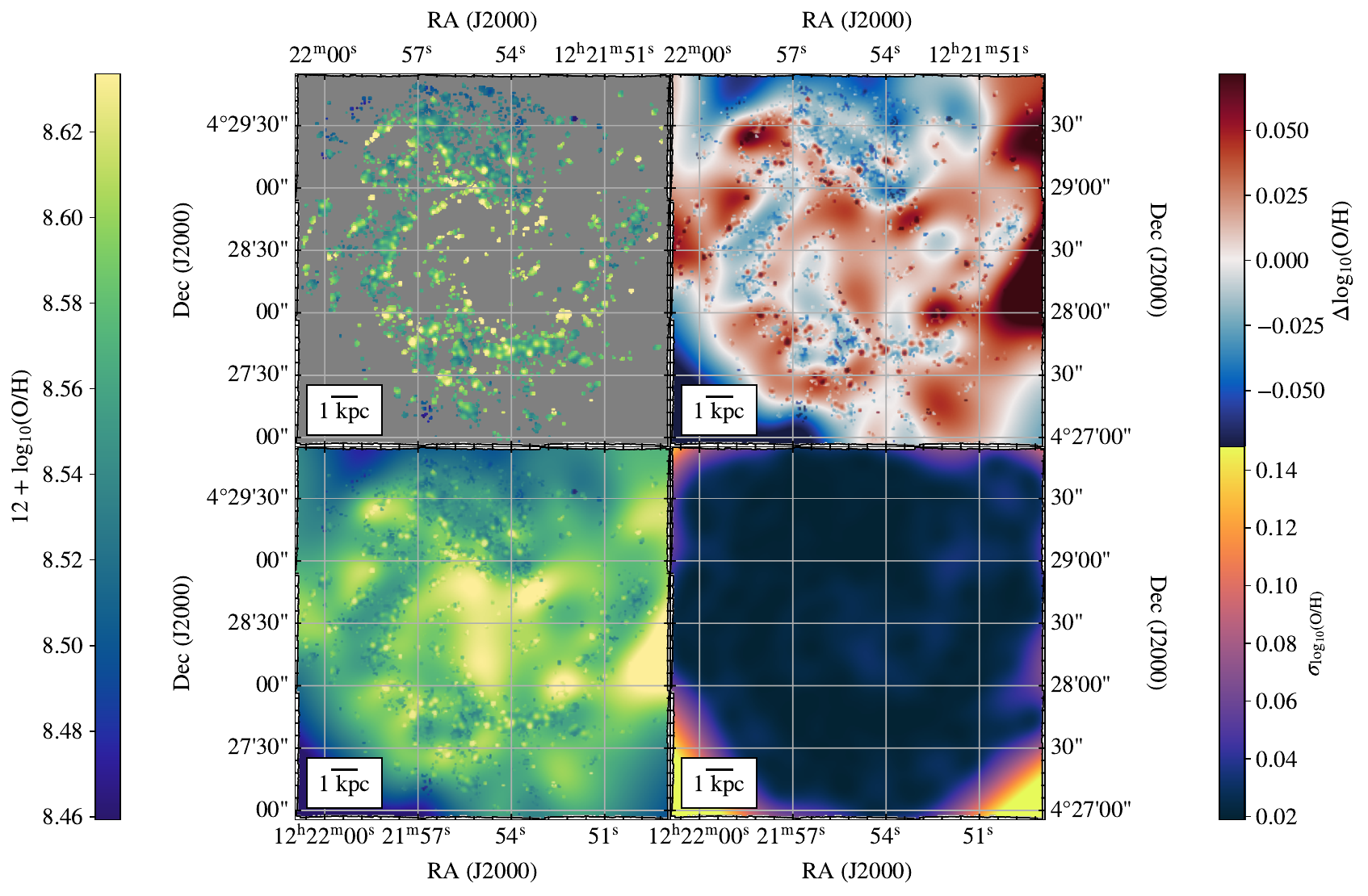}
\caption{As Fig. \ref{fig:gpr_fit}, but for NGC4303}
\end{figure*}
\newpage
\begin{figure*}
\includegraphics[width=2\columnwidth]{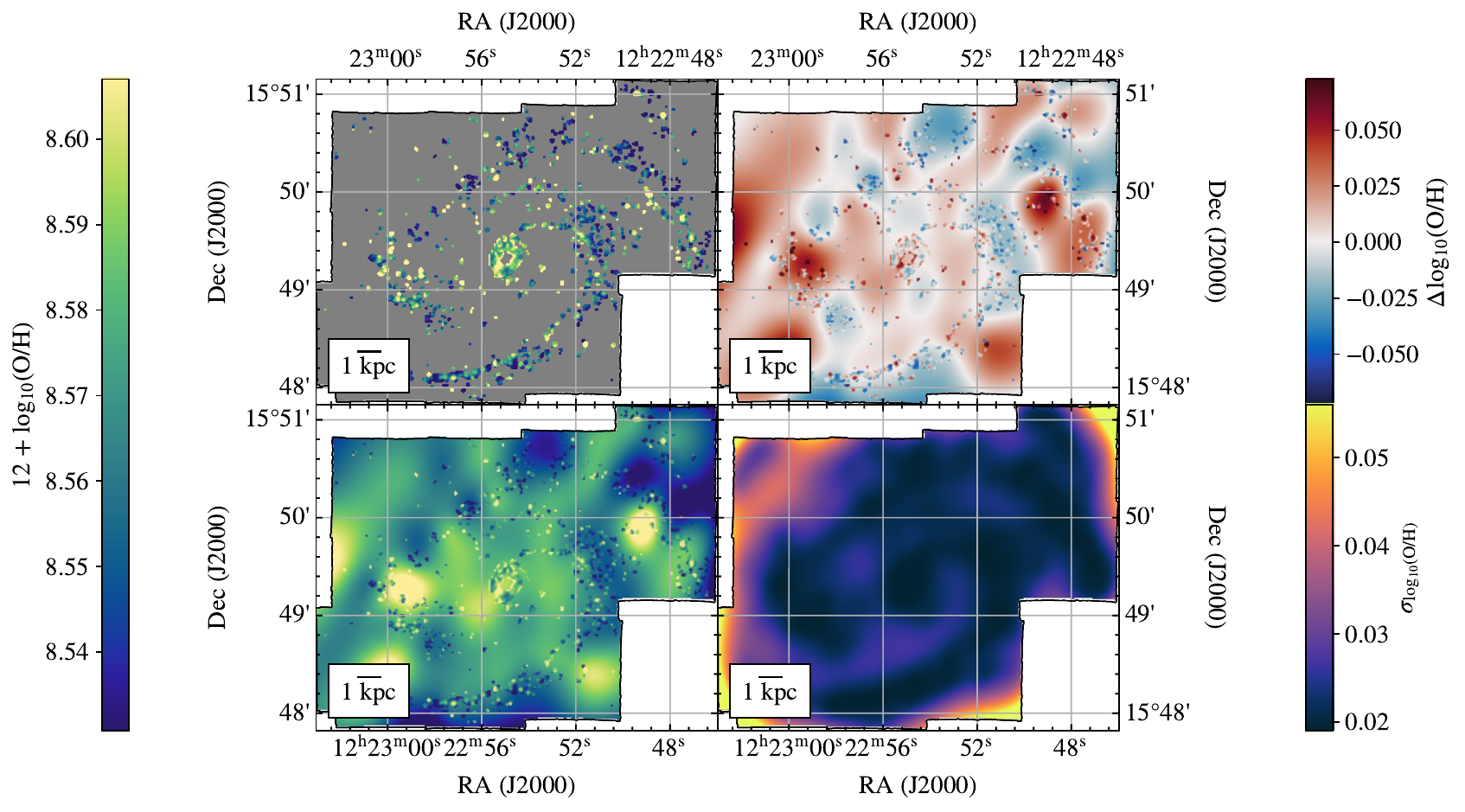}
\caption{As Fig. \ref{fig:gpr_fit}, but for NGC4321}
\end{figure*}
\newpage
\begin{figure*}
\includegraphics[width=2\columnwidth]{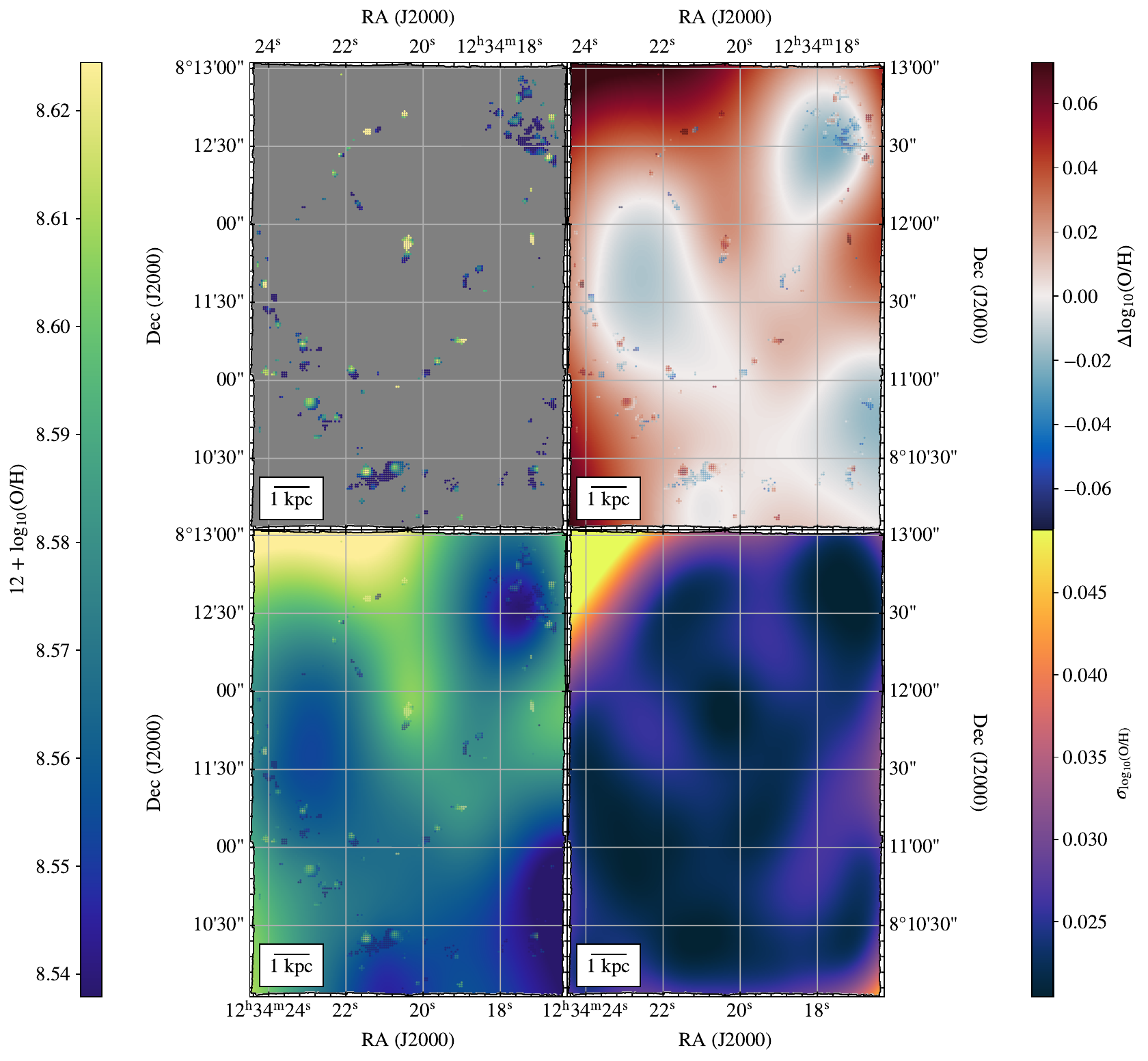}
\caption{As Fig. \ref{fig:gpr_fit}, but for NGC4535}
\end{figure*}
\newpage
\begin{figure*}
\includegraphics[width=2\columnwidth]{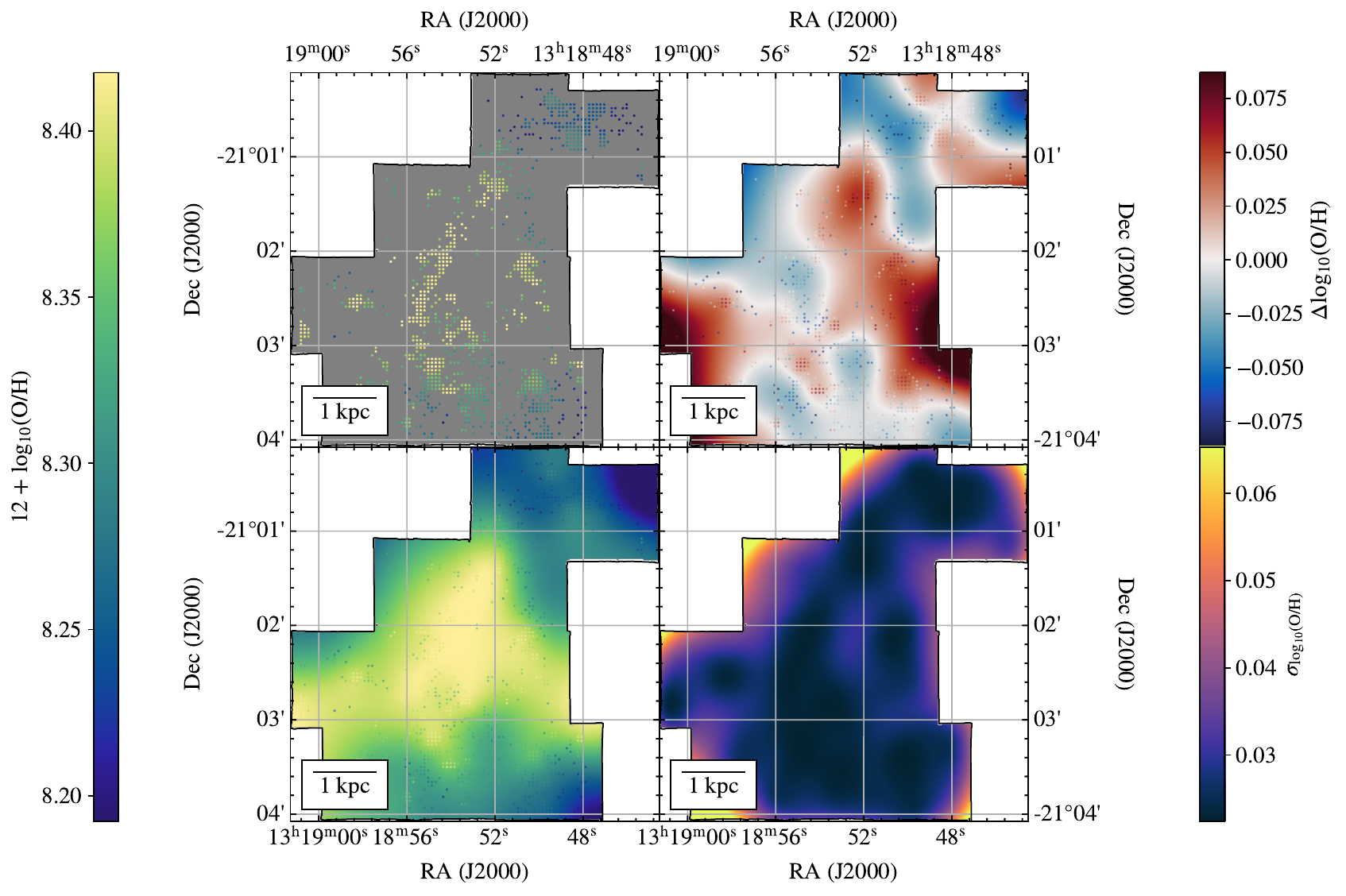}
\caption{As Fig. \ref{fig:gpr_fit}, but for NGC5068}
\end{figure*}
\newpage
\begin{figure*}
\includegraphics[width=2\columnwidth]{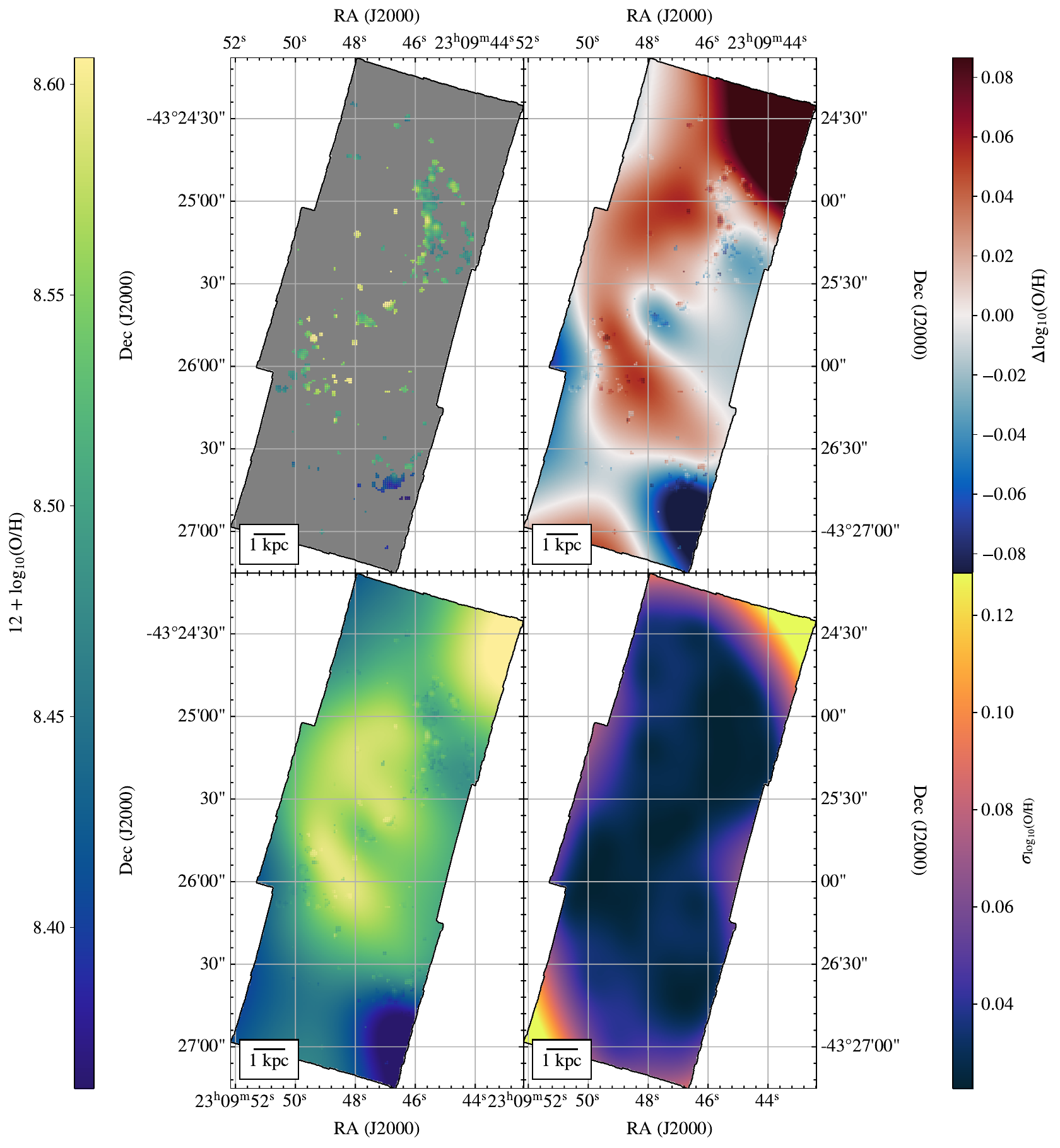}
\caption{As Fig. \ref{fig:gpr_fit}, but for NGC7496}
\end{figure*}
\newpage